\documentclass[%
aip,
amsmath,amssymb,
preprint,%
author-year,%
]{revtex4-1}

\usepackage{CJK}
\usepackage{graphicx}% Include figure files
\usepackage{dcolumn}% Align table columns on decimal point
\usepackage{bm}% bold math
\usepackage[mathlines]{lineno}% Enable numbering of text and display math
\usepackage[utf8]{inputenc}
\usepackage[T1]{fontenc}
\usepackage{mathptmx}
\usepackage{amsmath}

\begin{document}
%\linenumbers
\begin{CJK*}{UTF8}{}
\CJKfamily{gbsn}

\preprint{AIP/123-QED}

\title[]{The motion of respiratory droplets produced by coughing}
% Force line breaks with \\
\author{Hongping Wang (王洪平)}
\author{Zhaobin Li (\CJKfamily{bsmi}李曌斌)}
\author{Xinlei Zhang (张鑫磊)}
\author{Lixing Zhu (朱力行)}
\author{Yi Liu (刘毅)}
\author{Shizhao Wang (王士召)}%
\email{wangsz@lnm.imech.ac.cn}
\affiliation{The State Key Laboratory of Nonlinear Mechanics, Institute of Mechanics, Chinese Academy of Sciences, Beijing 100190, China%\\This line break forced with \textbackslash\textbackslash
}%

\date{\today}% It is always \today, today,

\begin{abstract}
Coronavirus disease 2019 (COVID-19) has become a global pandemic infectious respiratory disease with high mortality and infectiousness. This paper investigates respiratory droplet transmission, which is critical to understanding, modeling and controlling epidemics. In the present work, we implemented flow visualization, particle image velocimetry (PIV) and particle shadow tracking velocimetry (PSTV) to measure the velocity of the airflow and droplets involved in coughing and then constructed a physical model considering the evaporation effect to predict the motion of droplets under different weather conditions. The experimental results indicate that the convection velocity of cough airflow presents the relationship $t^{-0.7}$ with time; hence, the distance from the cougher increases by $t^{0.3}$ in the range of our measurement domain. Substituting these experimental results into the physical model reveals that the small droplets (initial diameter $D \leq$ 100 $\mu$m) evaporate to droplet nuclei and that the large droplets with $D \geq$ 500 $\mu$m and initial velocity $u_0 \geq$ 5 m/s travel more than 2 m. Winter conditions of low temperature and high relative humidity can cause more droplets to settle to the ground, which may be a possible driver of a second pandemic wave in the autumn and winter seasons.
\end{abstract}

\maketitle
\end{CJK*}

\begin{quotation}

\end{quotation}

\section{\label{sec:intro}Introduction}
Coronavirus disease 2019 (COVID-19), caused by the novel coronavirus SARS-CoV-2, has become a global pandemic infectious respiratory disease. According to the coronavirus resource center at Johns Hopkins University, more than 30 million people worldwide have been confirmed to have COVID-19, and there have been more than 1,000,000 deaths globally. Other infectious respiratory diseases such as flu, tuberculosis, severe acute respiratory syndrome (SARS) and Middle East respiratory syndrome (MERS) still threaten people's health and life. Understanding the transmission of pathogens is critical to modelling and controlling epidemics \cite[]{Bourouiba2014}. In general, virion-laden respiratory fluid droplets are released into the environment by infected people through expiratory events such as talking, coughing and sneezing. There are three possible transmission pathways between infectious and susceptible individuals \cite[]{Bourouiba2014,Huang2020,Mittal2020}: large droplet transmission, contact transmission and airborne transmission. In the first two pathways, healthy people can be infected by physically inhaling droplets or contacting contaminated surfaces within close proximity. These two pathways are termed direct short-range routes of pathogen transmission, as stated by \cite{Bourouiba2014}. Airborne transmission has been suggested to be an additional and important route for respiratory diseases \cite[]{Morawska2020,Liu2020}. The microdroplet (or droplet nuclei) generated via small droplet evaporation can travel a very long distance within ambient air \cite[]{Matthews1989,Dbouk2020}. \cite{Liu2020} investigated the aerodynamic nature of SARS-CoV-2 by measuring the viral ribose nucleic acid (RNA) in aerosols in different areas of two Wuhan hospitals during the outbreak of COVID-19. The aerosols concentrations of SARS-CoV-2 were much higher in the toilet used by patients with COVID-19 and in areas where crowds that included infected individuals gathered. They proposed that SARS-CoV-2 may be transmitted through aerosols.

Coughing, which can generate droplets and droplet nuclei, is a typical symptom of COVID-19. These virion-laden respiratory fluid droplets are expelled into the environment by powerful air flow during coughing. Measuring the air velocity, droplet number density, velocity and size distribution are critical to understanding and modeling virus transmission. \cite{Zhu2006} investigated the transport characteristics of saliva droplets produced by coughing in a calm indoor environment via particle image velocimetry (PIV) experiments and numerical simulation. The PIV results indicated that the initial velocity of cough airflow is as high as 22 m/s with an average velocity of 11.2 m/s. Numerical analyses that did not consider evaporation indicated that droplets smaller than 30 $\mu$m in diameter are mostly transported with the airflow, droplets with diameters of 50-200 $\mu$m are significantly affected by gravity, and droplets of 300 $\mu$m or larger are mostly affected by inertia. \cite{Gupta2009} measured the flow rate, flow direction and mouth opening area for 25 human subjects and found that the flow rate of a cough as a function of time can be defined as a combination of gamma-probability-distribution functions. \cite{Xie2009} carried out a series of experiments to measure the number and size of respiratory droplets produced by talking and coughing. The distribution of the droplet diameter at the origin was given in their paper. To predict the trajectory and settling speed of droplets, a discrete fallout model and a continuous fallout model were developed based on the theory of multiphase turbulent buoyant clouds by \cite{Bourouiba2014}. These models can be used with clinical data to yield better estimations of the range of airborne respiratory disease transmission. The influence of environmental conditions on droplet transmission was investigated using a three-dimensional numerical simulation by \cite{Dbouk2020}. It was found that saliva droplets can travel up to 6 m, which is much further than the 2 m social distance metric, when the wind speed varies from 4 km/h to 15 km/h.

Wearing face masks is a fundamental and efficient protection against virus transmission. The Schlieren optical method was applied to visualize cough airflow with and without standard surgical and N95 masks \cite[]{Tang2009}. The results indicated that human coughing produces a rapid turbulent jet into the surrounding air and that a mask can block the formation of this jet. \cite{David2012} used flow visualization to investigate the expelled air dispersion distances during coughing with a human patient simulator with and without a surgical mask or N95 mask in a negative pressure isolation room. The turbulent jet caused by coughing without a mask can propagate close to 0.7 m, while the air dispersion distance is approximately 0.15 m with an N95 mask. To clarify whether masks can offer effective protection against droplet infection, \cite{Kahler2020} conducted three experiments of fluid mechanics to analyze the blockage caused by masks, to determine the effectiveness of different filter materials and masks, and to verify the effects of leakage flows at the edges of masks. The effectiveness and leakage of face masks were qualitatively visualized by \cite{Verma2020} using smoke. Wearing masks can significantly curtail the speed and range of respiratory jets. \cite{Dbouk2020-1} used multiphase computational fluid dynamics in a fully coupled Eulerian-Lagrangian framework to investigate the droplet dynamics of mild coughing. This numerical results showed that a mask can reduce airborne droplet transmission. However, some droplets still spread around and away from the mask.

As stated by \cite{Asadi2020}, virologists and epidemiologists are racing to understand COVID-19 and how best to treat it. Many unknowns remain, but one thing is eminently clear: COVID-19 is both deadly and highly transmissible. In addition to direct large droplet transmission and contact transmission, airborne transmission is a potential important pathway, especially in an enclosed small space \cite[]{Asadi2020,Liu2020}. Therefore, investigation of the generation and evolution of expiratory activities such as breathing, talking, coughing and sneezing, which can produce virion-laden droplets and aerosols traveling with the airflow, is very significant \cite[]{Mittal2020}. Regarding coughing, which is the main symptom of COVID-19, most studies include qualitative visualization without dynamic analysis. In the present work, we use a high-speed PIV system to capture the spatial-temporal evolution of coughing \cite[]{HeGW2002,HeGW2009,HeGW2017}. We carried out three optics-based fluid experiments. First, we analyzed large-scale visualization images generated using cigarette smoke \cite{Gupta2009}. Second, we used the PIV technique to calculate the airflow velocity near the mouth and analyzed the time-resolved flow fields. Third, we used particle shadow velocimetry (\cite{Estevadeordal2005,Hessenkemper2018}, PSV) to track the trajectories of large droplets and analyzed the velocity distribution of the droplets. Finally, we constructed a model to consider the evaporation and motion of a droplet and analyzed the influence of weather conditions on droplet transmission. The rest of the paper is organized as follows. In Sec.~\ref{sec:method}, we first introduce the experimental facility, setup and data processing methods used in this work. In Sec. \ref{sec:results}, we present the results and discussion in accordance with these experiments. Finally, we offer conclusions in Sec.~\ref{sec:con}.

\section{Methodology}\label{sec:method}
\subsection{Experimental setup}\label{subsec:setup}
\begin{figure}
\begin{center}
		\includegraphics[width=0.45\textwidth]{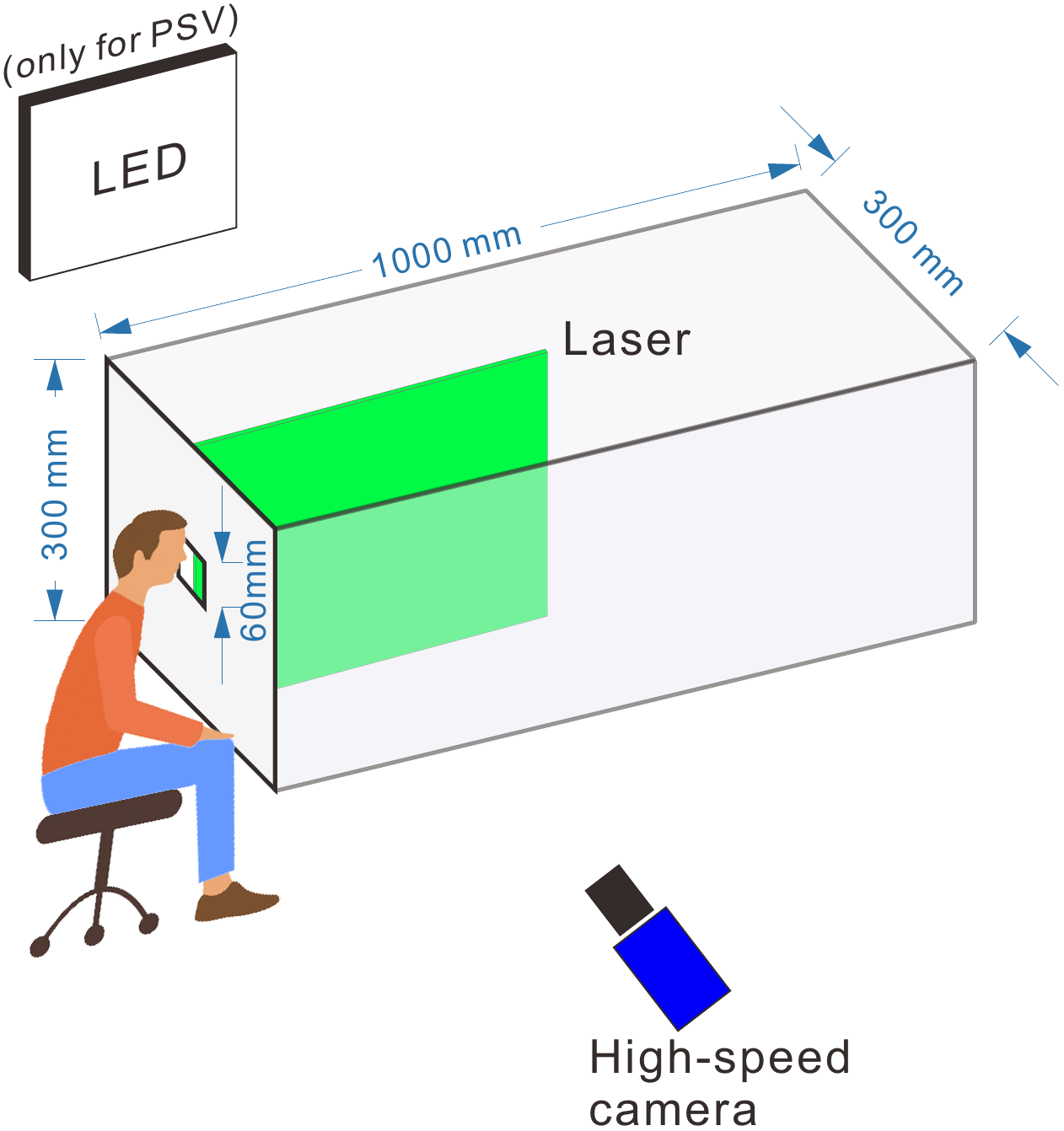}
\end{center}
\caption{Schematic diagram of the experimental setup.}\label{fig:setup}
\end{figure}
Figure \ref{fig:setup} shows a schematic diagram of the experimental setup, which consists of an acrylic tube, an LED plane, a laser generator and a high-speed camera. An opaque plate with a square-shared opening of 60 mm $\times$ 60 mm was mounted at the end of the tube to block the laser and protect the volunteers. The length, width and height of the tube were 1000 mm, 300 mm and 300 mm, respectively. When conducting the experiment, volunteers were asked to sit in front of the opaque plate and place their mouths as close as possible to the opening, and the exhaled airflow or droplets passed into the testing room through this opening. Additionally, volunteers were asked to wear safety goggles to protect their eyes. As stated in the introduction section, three kinds of experiments for different objectives were performed in this work. Each experiment was carried out by four healthy male volunteers, and each volunteer repeated three times. The detailed experimental configurations are listed in Tab. \ref{tab:exp_config} and are introduced in the next sections.

\begin{table}
	\caption{The experimental configurations for different cases.}\label{tab:exp_config} 
	\begin{ruledtabular}
		\begin{tabular}{p{1.2cm}<{\centering}p{2cm}<{\centering}p{2.5cm}<{\centering}p{2cm}<{\centering}p{2cm}<{\centering}p{2cm}<{\centering}}
			 & exposure time (sec) & sampling frequency (Hz) & image size (pix$^2$) & resolution (mm/pix) & light source\\
			\hline
			Visualization & 1/1024 & 1000 & 1280$\times$1024 & 0.25 & laser: 9 W\\
			PIV & 1/12584 & 10000 & 320$\times$256 & 0.25 & laser: 9 W \\
			PSV & 1/8000 & 5000 & 800$\times$400 &  0.035 & LED \\
		\end{tabular}
	\end{ruledtabular}
\end{table}

\subsection{Coughing visualization}\label{subsec:vis}
The coughing was visualized using cigarette smoke recorded by a high-speed Photron camera (FastcamSA2/86K-M3). The volunteer drew heavily on his cigarette and coughed out a cloud of smoke as naturally as possible. The diameters of cigarette smoke particles are mostly distributed in the submicron size range, i.e., 0.01--1 $\mu$m, as reported by \cite{Sahu2013}. A continuous laser sheet with a thickness of approximately 1 mm at a power of 9 W was used to illuminate the smoke. Images with a resolution of 1280$\times$1024 pixels$^2$ were recorded at 1000 frames per second (fps). A high-speed camera with a 50 mm F1.4 lens was placed approximately 1 m from the measurement domain, and this configuration yielded a field of view (FOV) of approximately 0.32$\times$0.26 m$^2$.

\subsection{Particle image velocimetry}\label{subsec:piv}
The velocity field near the mouth was estimated using the PIV technique. We still used a continuous laser sheet at a power of 9 W to illuminate the cigarette smoke exhaled during coughing. Different from the coughing visualization, the smoke images were acquired at a very high sampling frequency of 10000 Hz due to the high velocity of the flow near the mouth. Limited by the transfer bandwidth of the camera, the image resolution was reduced to 320$\times$256 pixels$^2$. The distance of the camera from the measurement domain was approximately 1.0 m, and a 50 mm F1.4 camera lens was used in this experiment. The size of the FOV was approximately 0.08$\times$0.06 m$^2$.

The velocity fields were estimated using in-house PIV software with a three-pass window deformation iterative multigrid scheme (WIDIM) \cite[]{Scarano2002}. Before the velocity estimation, the particle images were preprocessed using a Gaussian filter for image denoising. The interrogation window (IW) size used for the first pass was 48$\times$48 pixels$^2$, and the IW size of the final pass was set to 32$\times$32 pixels$^2$. The interval of the adjacent vectors was set to 2 pixels to increase the number of vectors. The cross-correlation map was calculated using a fast Fourier transform (FFT)-based approach, and the subpixel displacement was obtained by three-point Gaussian fitting. Outliers in the PIV fields were detected using the normalized median test proposed by \cite{Westerweel2005}, and the missing vectors were filled using linear interpolation. The final velocity fields were smoothed by average filtering with dimensions of 3$\times$3 to further reduce the measurement noise.

\subsection{Particle shadow tracking velocimetry}\label{subsec:PSV}
This experiment was performed to investigate the velocity of saliva droplets during coughing. To capture a clear droplet shape and track the trajectories, the PSV method was adopted. The flow region was illuminated using a high-power LED backlight, as shown in Fig. \ref{fig:setup}, instead of the laser sheet used in the smoke visualization and velocimetry. Additionally, a small depth of focus (DoF) corresponding to a small value of lens aperture was used to generate a thin volume where the droplets can be clearly identified. There were no droplets captured by our camera for dry coughing; therefore, the volunteer drank some water to moisten their throat and kept little water in their mouth. The remaining water in their mouth was driven by the pressure of coughing and generated many droplets due to hydrodynamic instability \cite[]{Scharfman2016,Mittal2020}. A camera with a 105 mm F2.4 lens was placed 0.4 m from the measurement domain. Images with a resolution of 800$\times$400 pixels$^2$ were recorded at a frequency of 5000 Hz, and the exposure time was 1/8000 seconds. The FOV size was 0.028$\times$0.014 m$^2$.

\begin{figure*}
\begin{center}
		\includegraphics[width=0.23\textwidth]{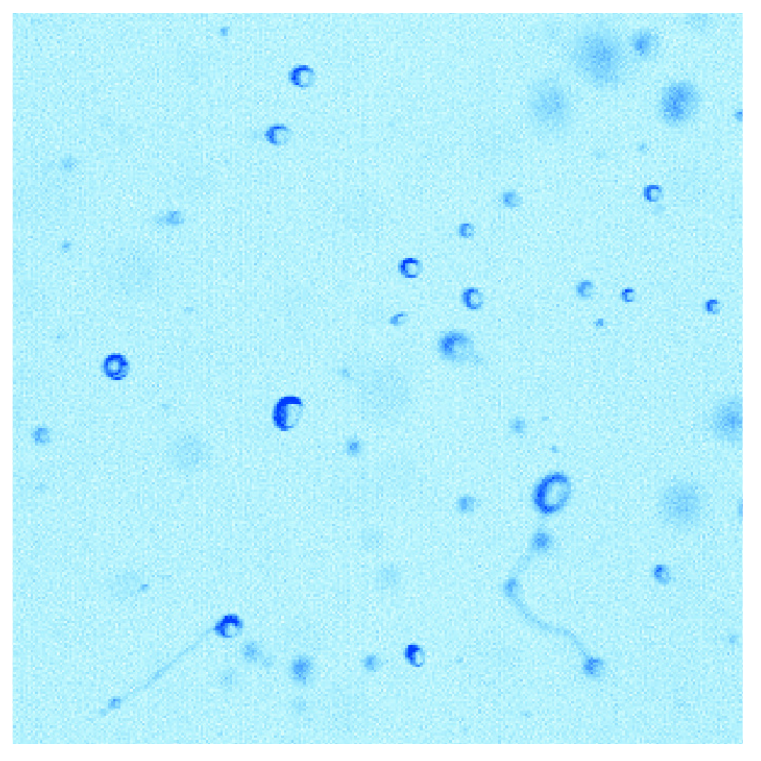}
\put(-100,93){(a)}
		\includegraphics[width=0.23\textwidth]{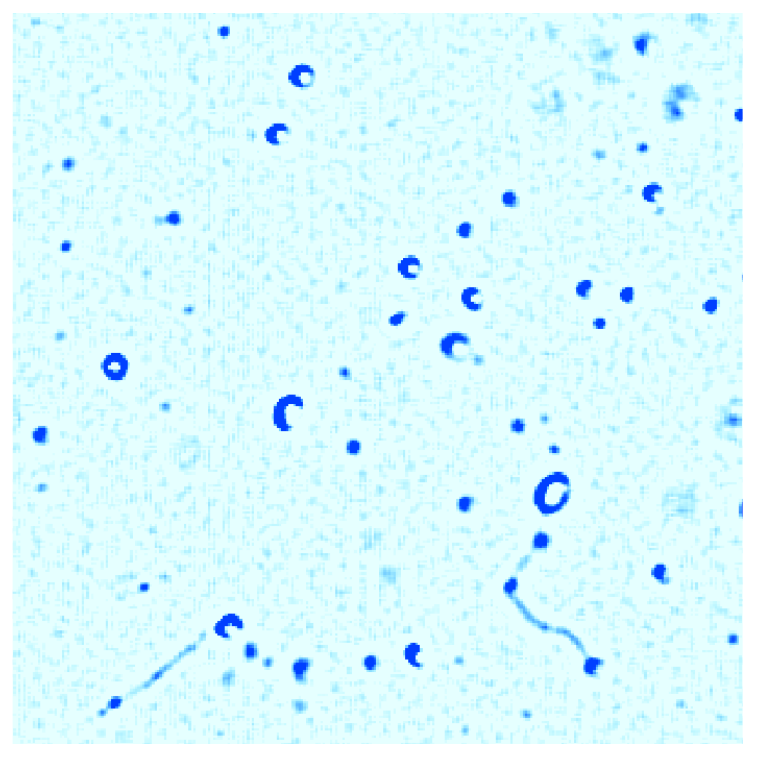}
\put(-100,93){(b)}
		\includegraphics[width=0.23\textwidth]{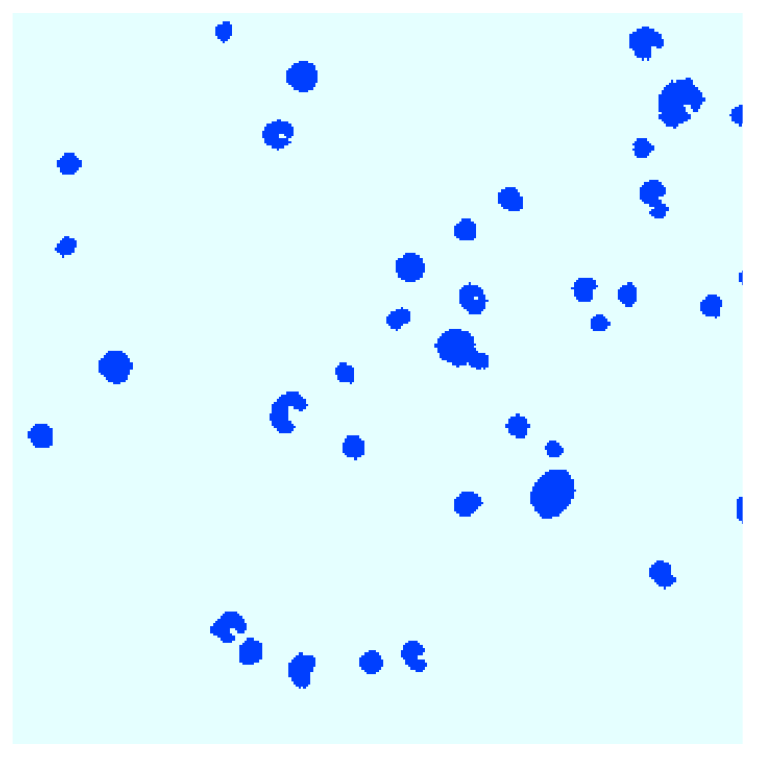}
\put(-100,93){(c)}
		\includegraphics[width=0.23\textwidth]{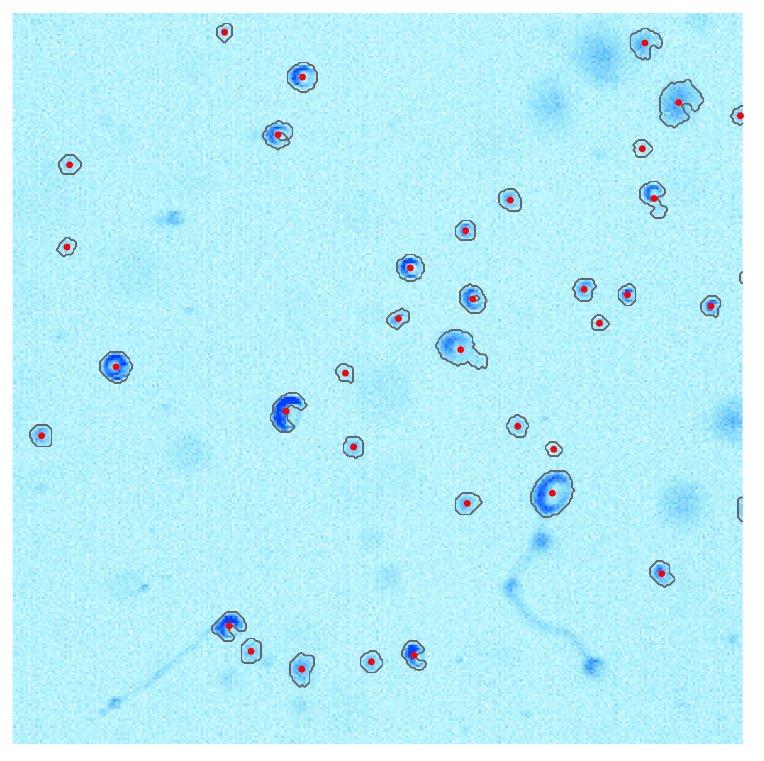}
\put(-100,93){(d)}
\end{center}
\caption{Illustrations of the droplet detection method used in the present work. (a) Part of the original image. (b) Image $L$ obtained from the Laplacian of the Gaussian method. (c) Binary image of the connected regions. (d) Detected droplets. The gray contour lines correspond to the edges of the droplets, and the red dots correspond to the locations.}\label{fig:psv_droplets}
\end{figure*}

Many droplets of different sizes were ejected from the volunteer's mouth during coughing, and only the droplets located in the region of the DoF were in focus in the images. To estimate a droplet's velocity, the location of the droplet had to be detected first. Segmentation methods based on a gray-level or gradient threshold are not appropriate for the droplets because of the intensity variation with the size of the droplets and the refraction of irregular shapes. Additionally, the presence of blurred droplets can locally modify the background around the in-focus droplets \cite[]{Castanet2013}. In the present work, the processes for droplet detection are given in Fig. \ref{fig:psv_droplets} as follows. First, the original image (Fig. \ref{fig:psv_droplets} (a)) was processed using the Laplacian of the Gaussian (LoG) method to obtain the intermediate image $L$ (Fig. \ref{fig:psv_droplets} (b)) as proposed by \cite{Castanet2013}. The droplet is in the region where the Laplacian is negative. The Gaussian filter size is 7$\times$7 pixels$^2$. Second, the connected regions of image $L$ were computed, corresponding to $L\le L_{max}$. To accurately determine the edges of the droplets, pixels with a negative Laplacian value ($-$0.5 was adopted in practice considering image noise) were connected with the region of $L\le L_{max}$ to generate a complete droplet. The value of $L_{max}$ was set to $-10$ in this study, and then regions that were too large or too small were excluded. Figure \ref{fig:psv_droplets} (c) shows the resulting region of connection. Third, a contour operation was applied to the binary image of the connected regions to find the edge of the droplets \cite[]{Castanet2013}, and the geometric center of this region was calculated as the location of the droplet. As shown in Fig. \ref{fig:psv_droplets} (d), the gray contour lines present good agreement with the edges of droplets, and the red dots represent the locations of the droplets. Note that a few droplets present a crescent shape due to light refraction.

\begin{figure}
\begin{center}
		\includegraphics[width=0.3\textwidth]{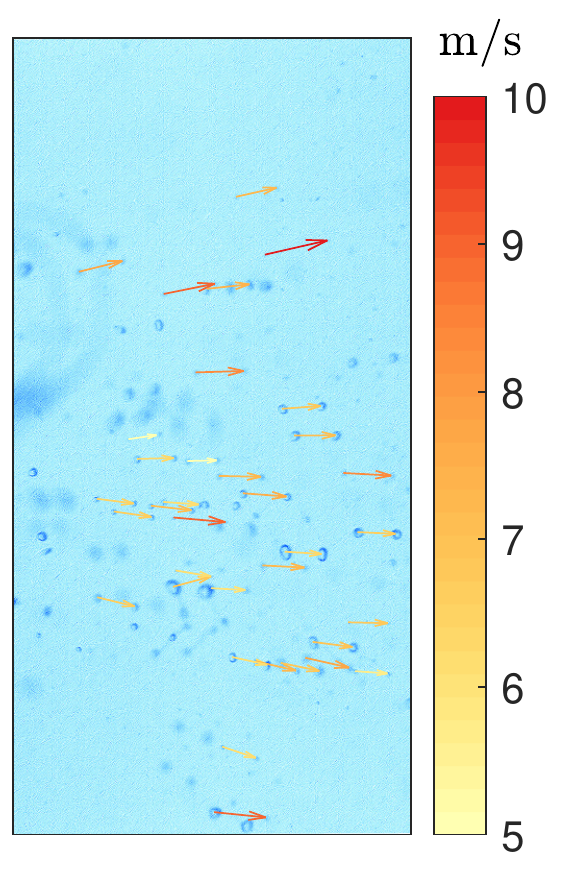}
\end{center}
\caption{Example of the velocity vectors of expelled droplets.}\label{fig:psv_velocity}
\end{figure}

After droplet detection, the velocity of the droplets is determined using the noniterative double-frame particle tracking velocimetry (PTV) proposed by \cite{Fuchs2017}. The combination of particle shadow images and PTV is referred to as particle shadow tracking velocimetry (PSTV) \cite[]{Hessenkemper2018}. We did not track the trajectories of the droplets because the measurement domain in the $x$ direction is as short as 400 pixels. Figure \ref{fig:psv_velocity} gives an example of the velocity vectors of the droplets obtained using PSTV; the minimum and maximum droplet velocities are 5.0 m/s and 10.0 m/s, respectively.

\section{Results and Discussion} \label{sec:results}
\subsection{Convection velocity from visualization}\label{subsec:conection}
\begin{figure*}
\begin{center}
		\includegraphics[height=0.228\textwidth]{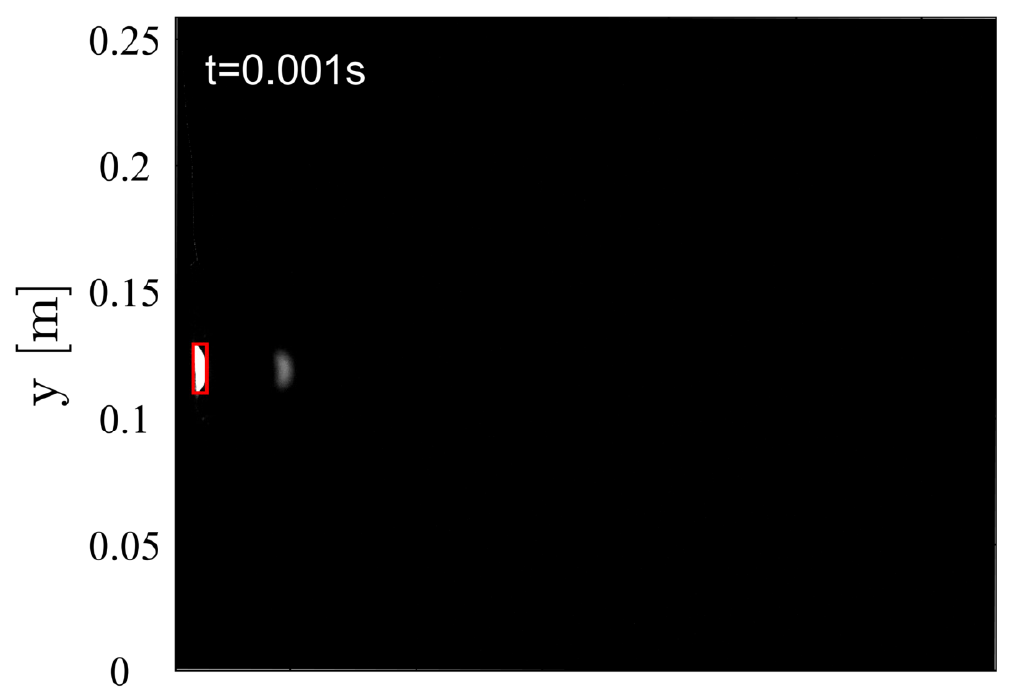}
		\includegraphics[height=0.228\textwidth]{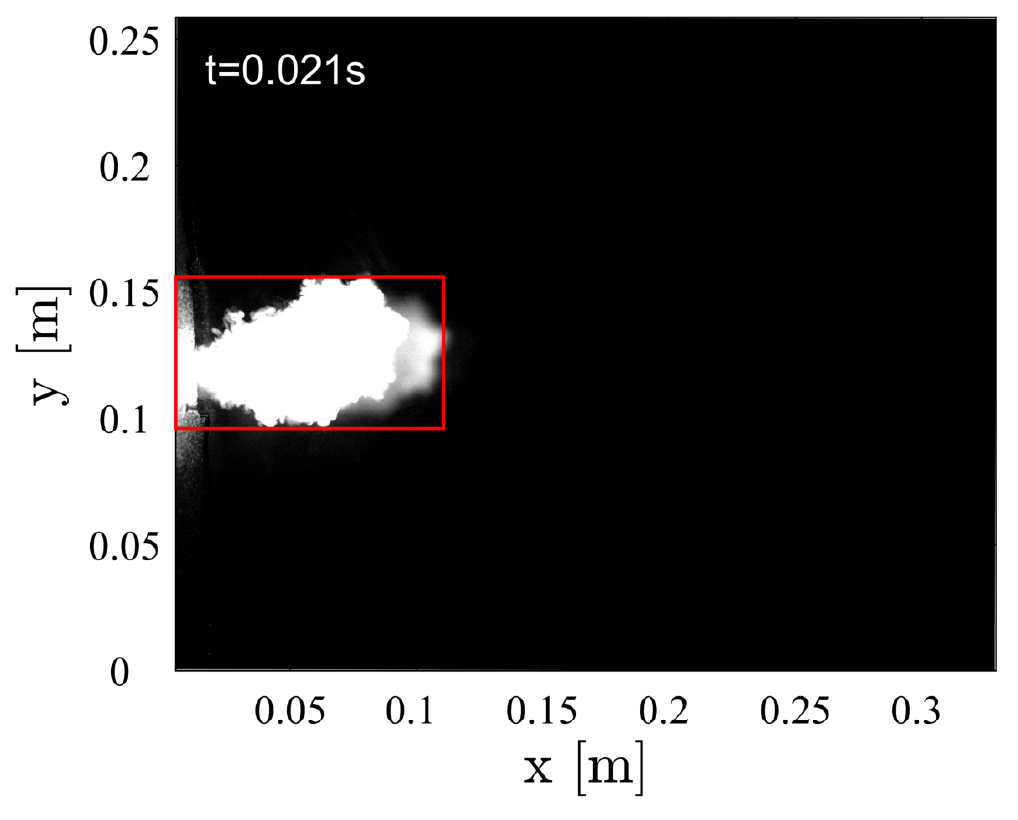}
		\includegraphics[height=0.228\textwidth]{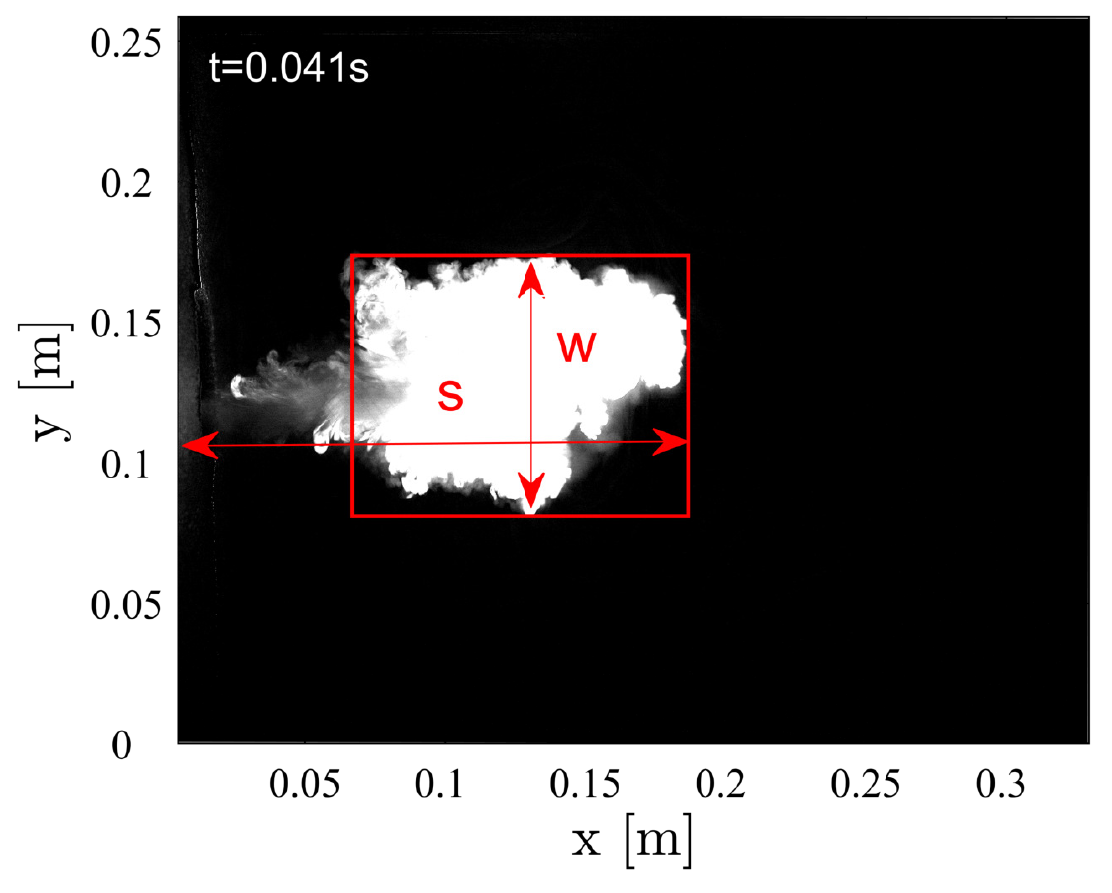}
\\
		\includegraphics[height=0.27\textwidth]{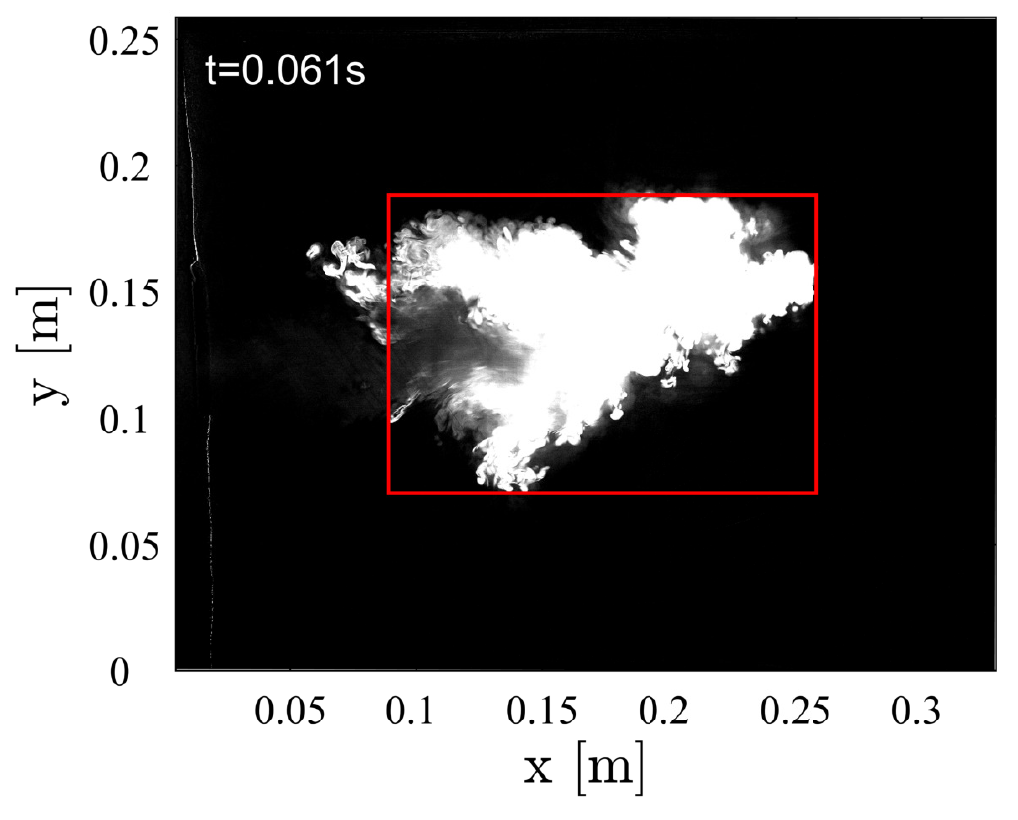}
		\includegraphics[height=0.27\textwidth]{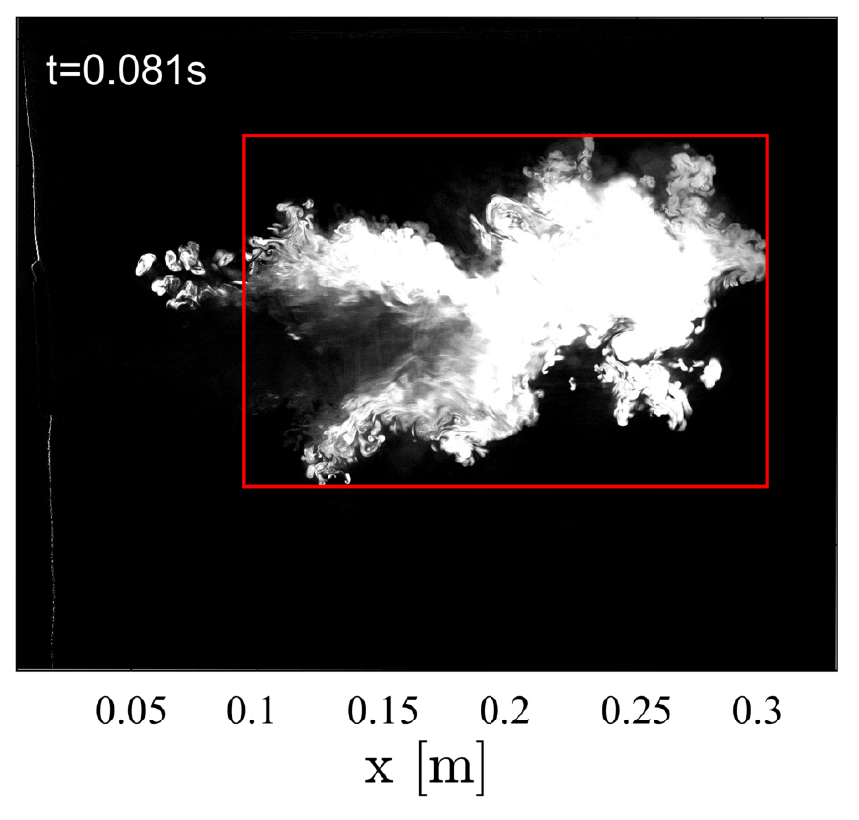}
		\includegraphics[height=0.27\textwidth]{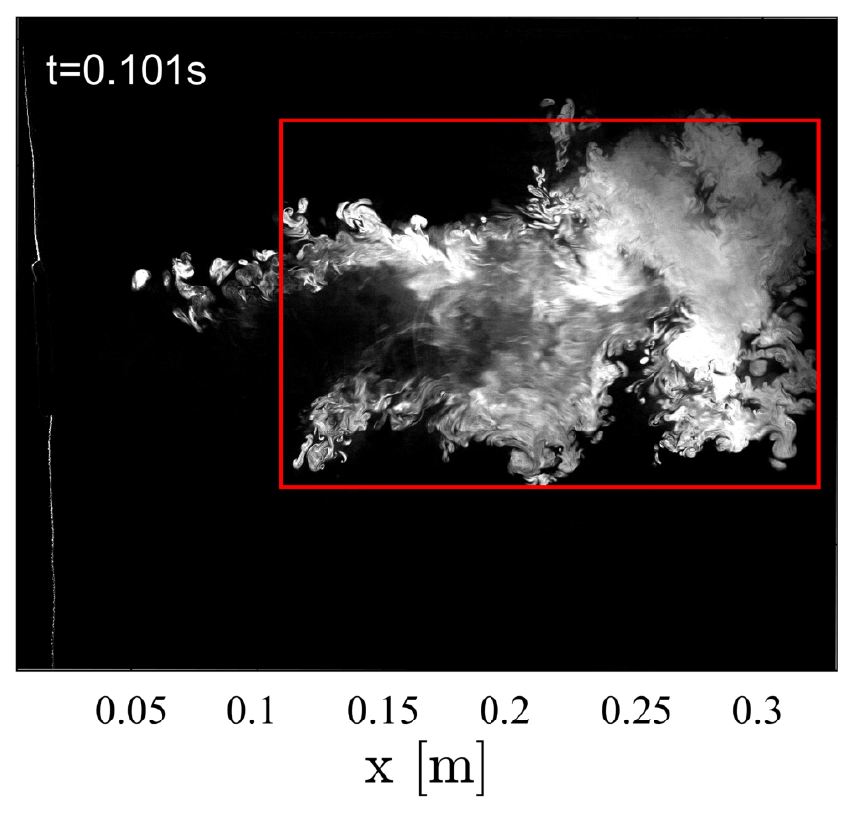}
\end{center}
\caption{Convection of smoke expelled by coughing. The image sequence shows a cough at intervals of 0.02 s. The red rectangles indicate the regions with higher gray levels, corresponding to concentrated smoke. The distance $s$ is defined as the distance from the left origin to the right border of the rectangle, and $w$ is the width of the box, as shown in the image of t=0.041 s. }\label{fig:visual_gray}
\end{figure*}

The coughing process is visualized by the expelling of smoke. Figure \ref{fig:visual_gray} shows the evolution of a cough at a time interval of 0.02 s. The onset of coughing is determined by visually inspecting the image sequence. The red rectangles indicate regions with high gray levels, corresponding to concentrated smoke. The distance $s$ is defined as the distance from the left origin to the right border of the rectangle, and $w$ denotes the width of the rectangles. At the beginning of coughing, the ejected smoke presents a cone-like shape and a high concentration and then rapidly transitions into turbulence. With the entraining of the ambient quiescent air, the cough smoke is gradually decelerated and reaches the right border of the FOV (approximately 0.3 m) at approximately 0.1 s.

\begin{figure}
\begin{center}
		\includegraphics[height=0.3\textwidth]{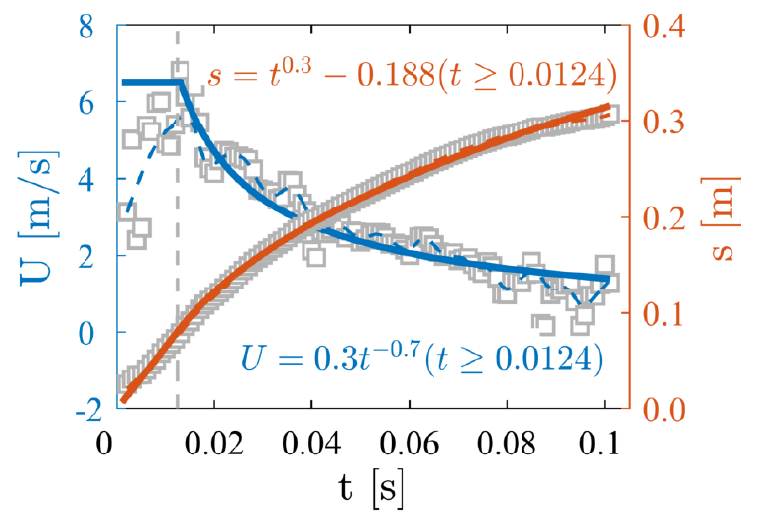}
		\quad
		\includegraphics[height=0.3\textwidth]{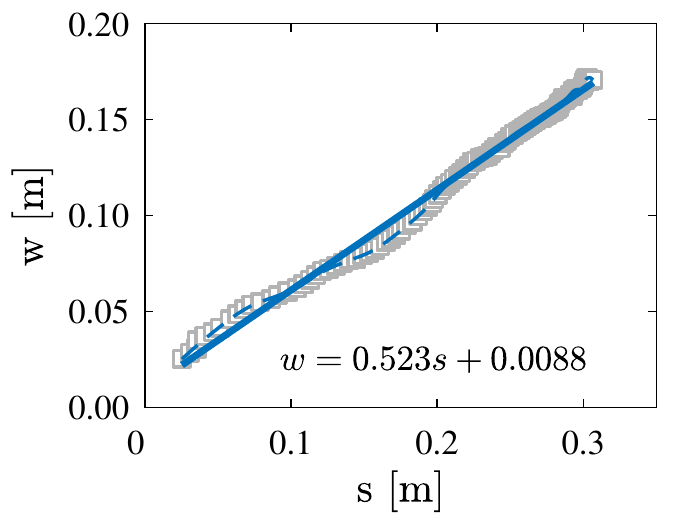}
		\put(-300,-9){(a)}
		\put(-85,-9){(b)}
\end{center}
\caption{(a) Convection velocity $U_c$ and distance $s$ as a function of time $t$. (b) Relationship between the distance $s$ and the width $w$ of the cough jet. \textit{Open gray squares}: the original data averaged over 11 cases. \textit{Dashed lines}: the smoothed results. \textit{Solid lines}: the fitting results corresponding to the given formulas.}\label{fig:visual_fit}
\end{figure}

The convection velocity $U_c$ and distance $s$ as a function of time $t$ averaged over 11 cases (one case is excluded because of unreliable data) are presented in Fig. \ref{fig:visual_fit}. In this work, the convection velocity is defined as $U_c = ds/dt$ and computed using a central difference. The open gray squares, dashed lines and solid lines represent the original experimental data, smoothed data and fitted data, respectively. The convection velocity $U_c$ presents a peak value of approximately 6 m/s at $t_p$ = 0.0124 s. To fit the curve of $s$, we consider $U_c$ to be a piecewise function: $U_c$ holds a constant value of 6.48 m/s for $t < t_p$ and then decreases as $U_c = 0.3t^{-0.7}$ for $t\ge t_p$. The distance $s$ can be integrated as $s = \int_{0}^{t}U_cdt$; hence, $s \sim t^{0.3}$ at $t \geq t_p$. This result is different from that of \cite{Bourouiba2014}, who stated that $s$ is related to $t$ by $s \sim t^{0.5}$ in the first phase, which is dominated by jet-like dynamics. We believe that $s$ varies as $t^{0.3}$ because a cough is a pluse-like jet and not a continuous jet. In accordance with the fitting functions, the relationship between $U_c$ and $s$ can be expressed as:
\begin{equation}\label{eq:uc_s}
	U_c = \left\{ 
	\begin{split}
		& 6.48 \quad m/s, 				&s \leq 0.08 \quad m,\\
		& 0.3(s+0.188)^{-7/3} \quad m/s,       &s \ge 0.08 \quad m.
	\end{split}
	\right.
\end{equation}
This equation is used to model the motion of cough airflow in Sec. \ref{subsec:model}. If we do not consider the temperature of the exhaled airflow and we assume that the indoor airflow velocity is approximately 0.3 m/s on average \cite[]{Gong2006}, the airflow exhaled by coughing can be calculated to travel approximately 0.8 m from the mouth by substituting 0.3 m/s into the formulas in Fig. \ref{fig:visual_fit}. Wearing masks and maintaining social distance are important for preventing airborne transmission.

The dependence of the width $w$ on the distance $s$ is illustrated in Fig. \ref{fig:visual_fit} (b). It can be seen from this figure that $w$ increases linearly as $s$ increases. The fitted function is written as $w = \alpha s+\beta$, where $\alpha = 0.523$ and $\beta = 0.0088$. The $\alpha$ reported by \cite{Bourouiba2014} is 0.22, which is much smaller than the value obtained from our experiments. The linear relationship between $s$ and $w$ implies the self-similar growth of the cough jet.

\subsection{Flow dynamics of the airflow}
\begin{figure*}
\begin{center}
		\includegraphics[height=0.3\textwidth]{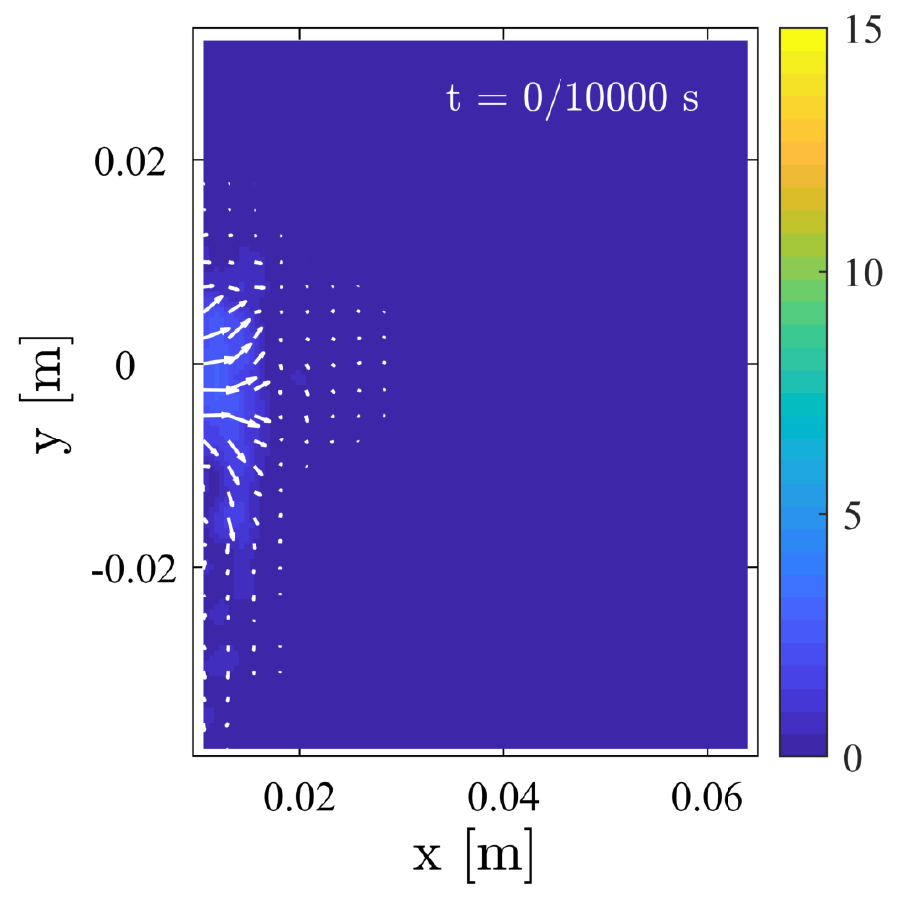}
		\includegraphics[height=0.3\textwidth]{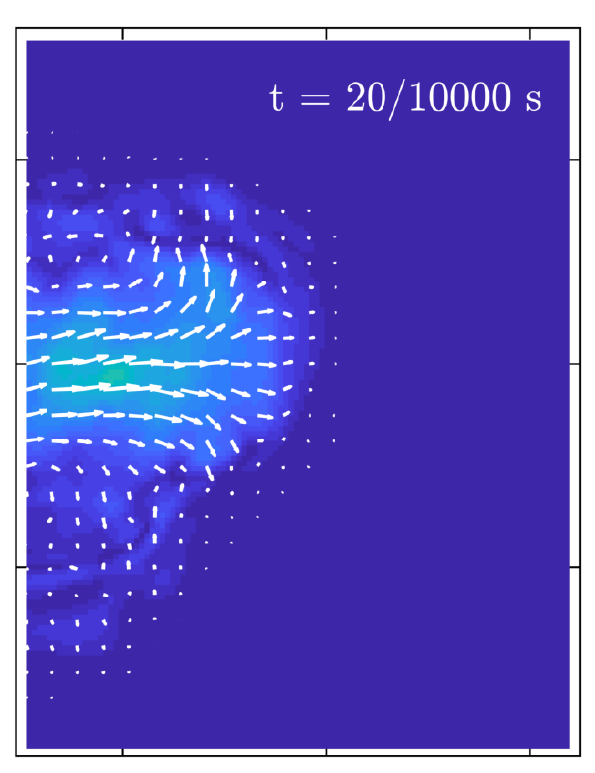}
		\includegraphics[height=0.3\textwidth]{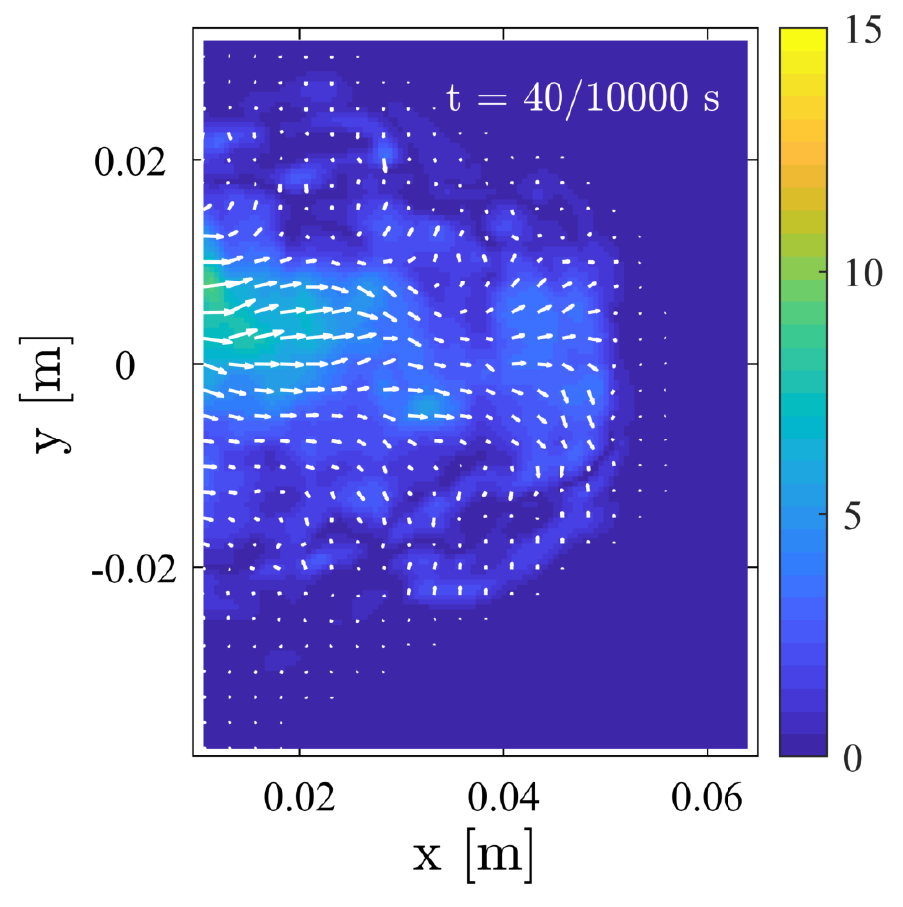}
\\
		\includegraphics[height=0.352\textwidth]{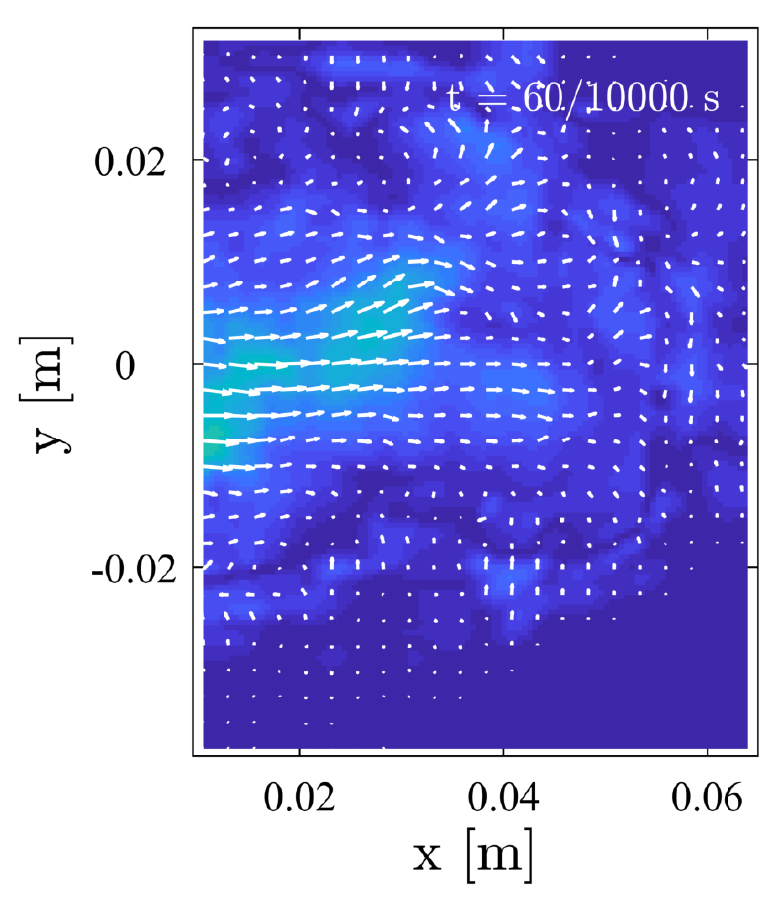}
		\includegraphics[height=0.352\textwidth]{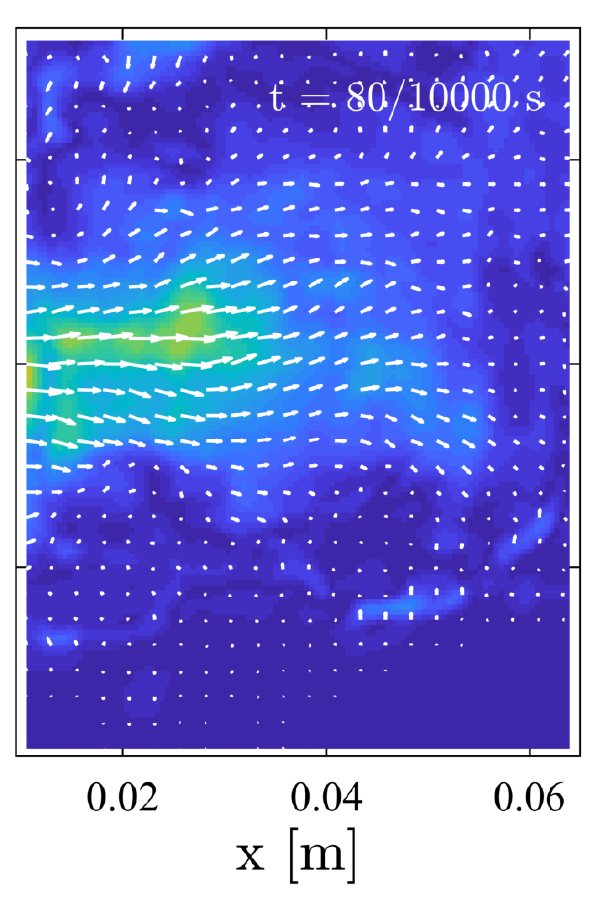}
		\includegraphics[height=0.352\textwidth]{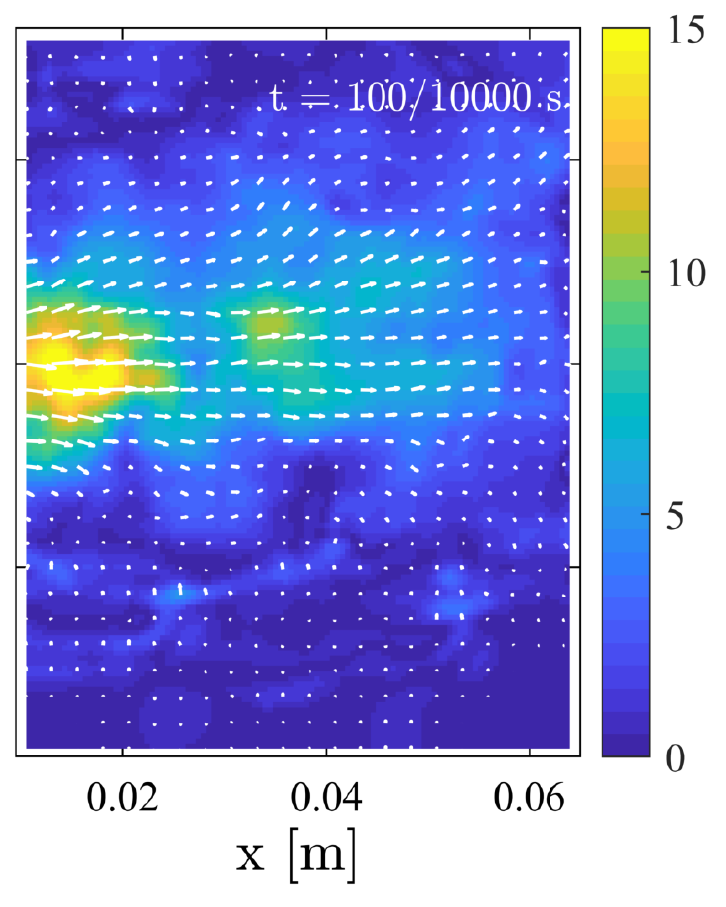}
\end{center}
\caption{Average velocity fields from time $t$ = 0 s to 0.01 s at intervals of 0.002 s.}\label{fig:airflow_inst}
\end{figure*}

The near-mouth velocity fields were estimated from the smoke cough using PIV. All 12 cases are averaged corresponding to the time to remove outliers and noise. The time $t$ = 0 s is assigned to when the smoke has just moved into the measurement domain. Only 800 velocity fields (approximately 0.08 s) are considered because the smoke gradually diffuses to a low gray level. Figure \ref{fig:airflow_inst} shows the averaged velocity fields from $t$ = 0 s to 0.01 s at a time interval of 0.002 s. The origin of the coordinates is adjusted in accordance with the position of the mouth. At $t$ = 0 s, the velocity of released smoke $u_0$ is approximately 4.0 m/s. \cite{Gupta2009} experimentally estimated that the mouth opening area of male subjects is approximately 4.00$\pm$0.95 cm$^2$. Therefore, the Reynolds number $Re = u_0d/\nu$ is approximately 5757, where $d$ = 0.023 m represents the diameter of the mouth when coughing and $\nu$ is the kinematic viscosity of air at a temperature of 30$^{\circ}$C. From the visual inspection of the velocity fields and the smoke images, the axis evolution of the cough jet presents strong turbulence characteristics due to entrainment and diffusion.

\begin{figure*}
\begin{center}
		\includegraphics[height=0.35\textwidth]{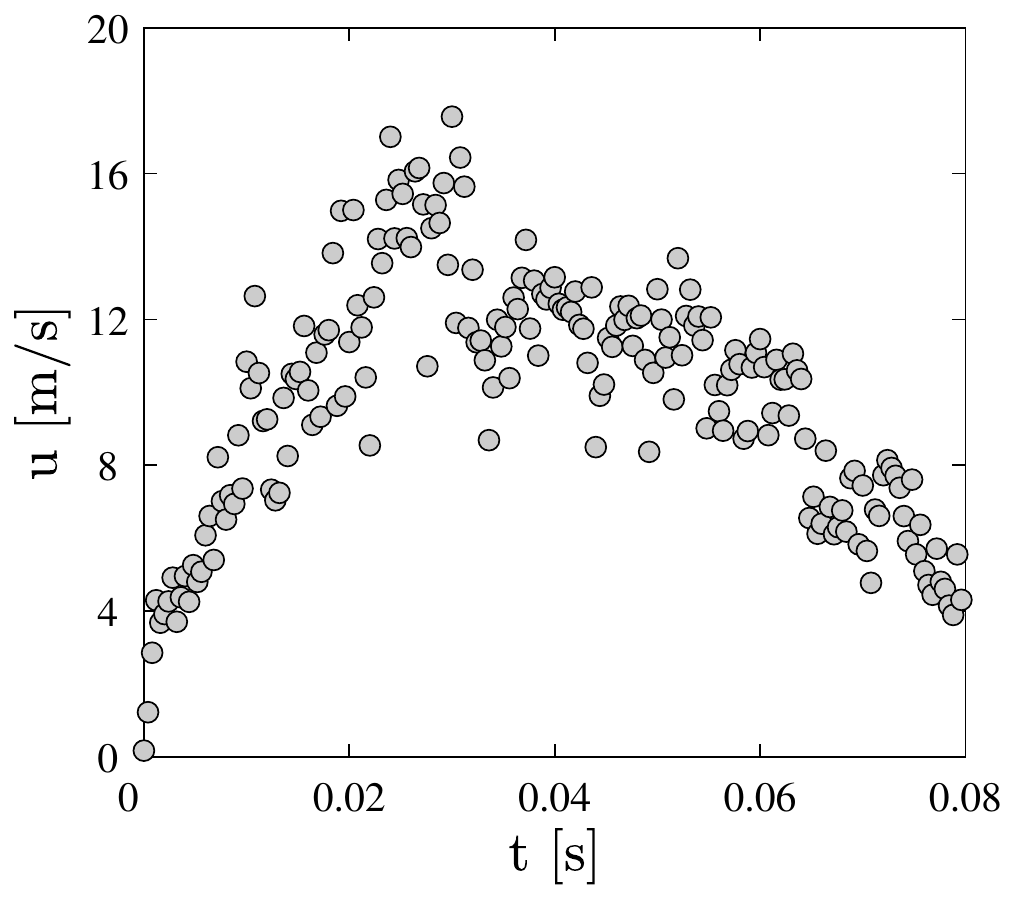}
		\includegraphics[height=0.35\textwidth]{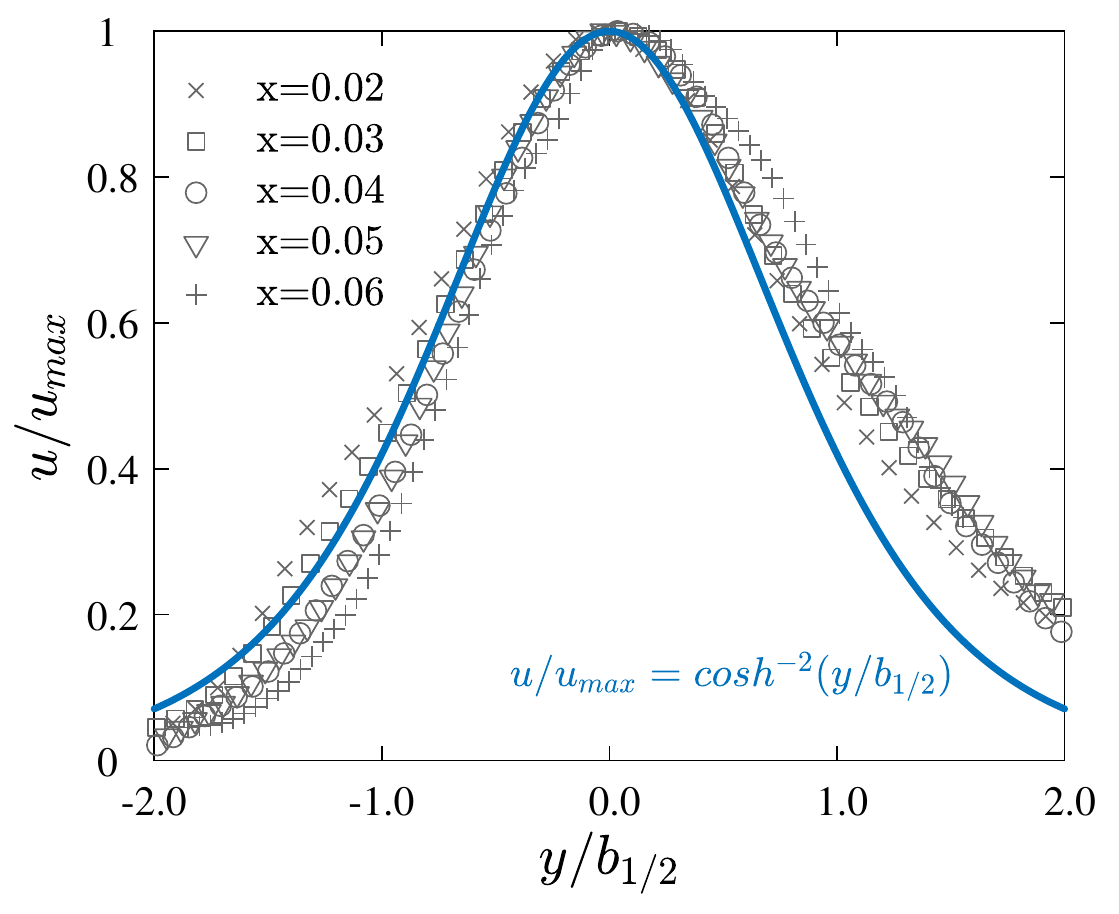}
		\put(-295,-9){(a)}
		\put(-105,-9){(b)}
\end{center}
\caption{(a) Velocity at position ($x$, $y$) = (0.02, 0) as a function of time. (b) Normalized axial mean velocity profiles at different streamwise positions of $x$ = 0.02, 0.03, 0.04, 0.05 and 0.06 m.}\label{fig:airflow_norm_vel}
\end{figure*}

The velocity at position ($x$, $y$) = (0.02, 0) as a function of time is presented in Fig. \ref{fig:airflow_norm_vel} (a). The initial velocity at $t$ = 0 s is approximately 4.0 m/s and then rapidly increases to 15.0 m/s at $t$ = 0.03 s. The velocity gradually reduces to 4.0 m/s at time $t$ = 0.08 s. The variation in the velocity qualitatively agrees with the results given by \cite{Gupta2009}, \cite{Dudalski2020} and \cite{Kwon2012}. However, \cite{Dudalski2020} reported that the peak velocity value is approximately 22.0 m/s at $t$ = 0.068 s, which is different from the present result. A potential reason for the difference is that the position of the mouth and the onset time of coughing are slightly different among the volunteers in our experiments. These uncertainties are difficult to avoid under the current experimental conditions.

Figure \ref{fig:airflow_norm_vel} (b) shows the dimensionless time-averaged axial velocity at positions of $x$ = 0.02, 0.03, 0.04, 0.05 and 0.06 m. The mean velocity is normalized using the local maximum $u_{max}$ at position $x$, and the radial position is normalized by the jet half-width $b_{1/2}$, which is defined as the radial distance corresponding to half of $u_{max}$ \cite[]{Krishnan2010,Xu2013,Xu2018}. In this figure, a hyperbolic cosine function $u/u_{max}=cosh^{-2}(y/b_{1/2})$ that is used to describe the velocity profiles of synthetic jets \cite[]{Zhang2007,Xu2013,Xu2018} is given for comparison. The normalized velocity profiles agree well with the blue curve when $y/b_{1/2}\leq0$, while velocity when $y/b_{1/2} > 0$ is higher than the blue curve. This asymmetric axial velocity profile may be due to two possible reasons. First, the temperature of the expelled airflow is higher than that of the outside environment, which can cause the airflow to be driven by buoyancy \cite[]{Bourouiba2014}. In our opinion, this is the primary reason. Second, the number of samples is still not high enough to obtain converged results.

\begin{figure*}
\begin{center}
		\includegraphics[width=0.45\textwidth]{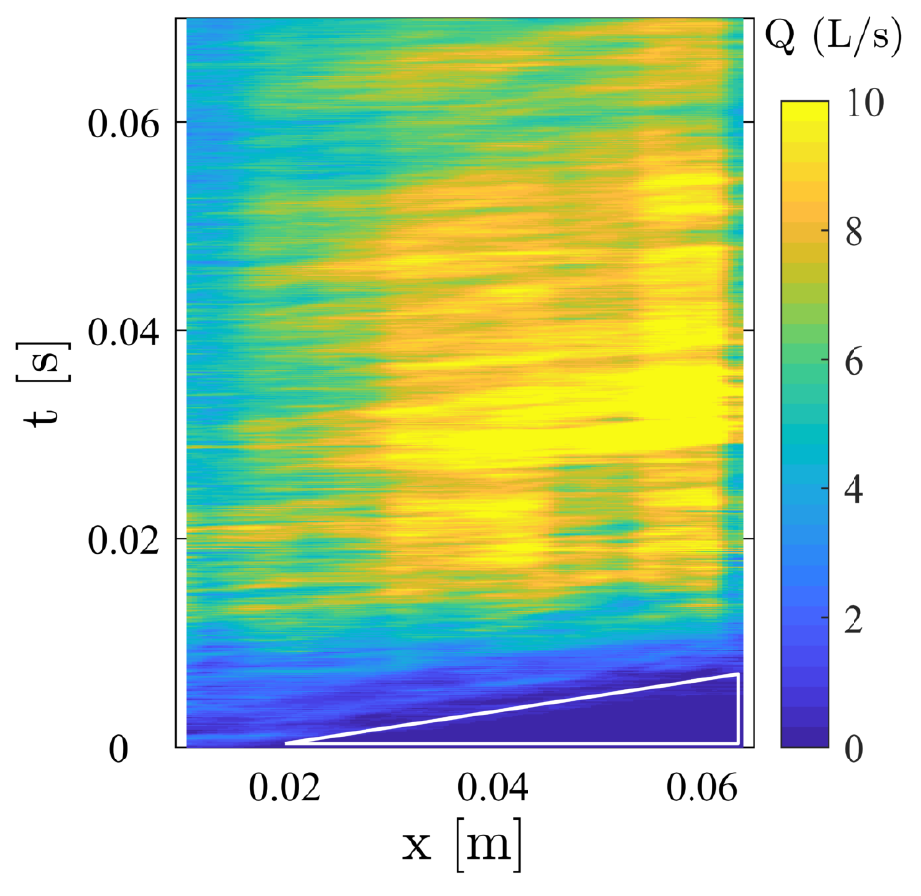}
		\includegraphics[width=0.45\textwidth]{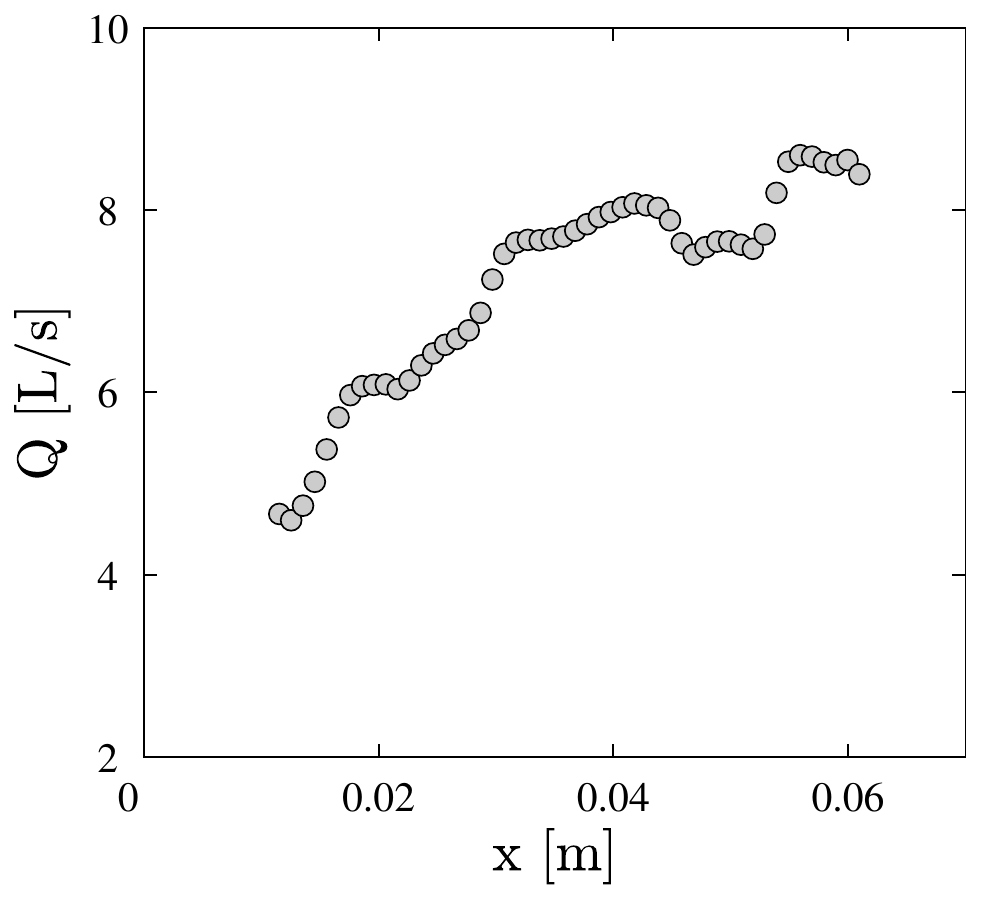}
		\put(-325,-9){(a)}
		\put(-105,-9){(b)}
\end{center}
\caption{(a) Flow rate $Q$ as a function of time $t$ and axial position $x$. (b) Time-averaged axial evolution of the flow rate $Q$.}\label{fig:airflow_Q}
\end{figure*}

The cough jet interacts with the surrounding air via flow entrainment, which can be viewed as a combination of small-scale nibbling plus a large-scale engulfment and induced inflow \cite[]{Philip2012}. The entrainment of a cough is qualitatively stronger than that of a continuous turbulent jet or synthetic jet because of the variation in the mouth position and flow direction. Figure \ref{fig:airflow_Q} (a) shows the flow rate $Q$ as a function of time $t$ and axial position $x$. The flow rate $Q$ is roughly estimated as:
\begin{equation}\label{eq:01}
Q = \pi\int_{-\infty}^{+\infty}u|r|dr,	
\end{equation}
where we assume that the cough jet is axisymmetric and $|r|$ is the absolute value of the radial location. The white triangle in the lower right of this figure indicates that the jet spreads from $x$ = 0.02 m to $x$ = 0.0635 m within 0.007 s, implying that the initial convection velocity is approximately 6.2 m/s. This value agrees with the result in Fig.~\ref{fig:visual_fit} obtained using flow visualization (6.48 m/s). Taking the axial location $x$ = 0.04 m as an example, the flow rate $Q$ first increases and then decreases with time $t$. This trend qualitatively agrees with the typical flow generated from a cough over time given by \cite{Gupta2009}. The time-averaged $Q$ from $t$ = 0.01 s to 0.07 s is presented in Fig. \ref{fig:airflow_Q} (b). It is obvious that the flow rate $Q$ increases with the axial position. The entrainment can be estimated from the axial derivative of $Q$ as ${dQ}/{dx}$, and the increase in $Q$ implies that the jet can entrain the ambient air. However, it is difficult to quantize %Editor: Please consider replacing the preceding text with 'quantify'.
the entrainment effect of coughing in the limited measurement domain of the data.

\subsection{Velocity of the large droplets}
\begin{figure*}
\begin{center}
		\includegraphics[height=0.3\textwidth]{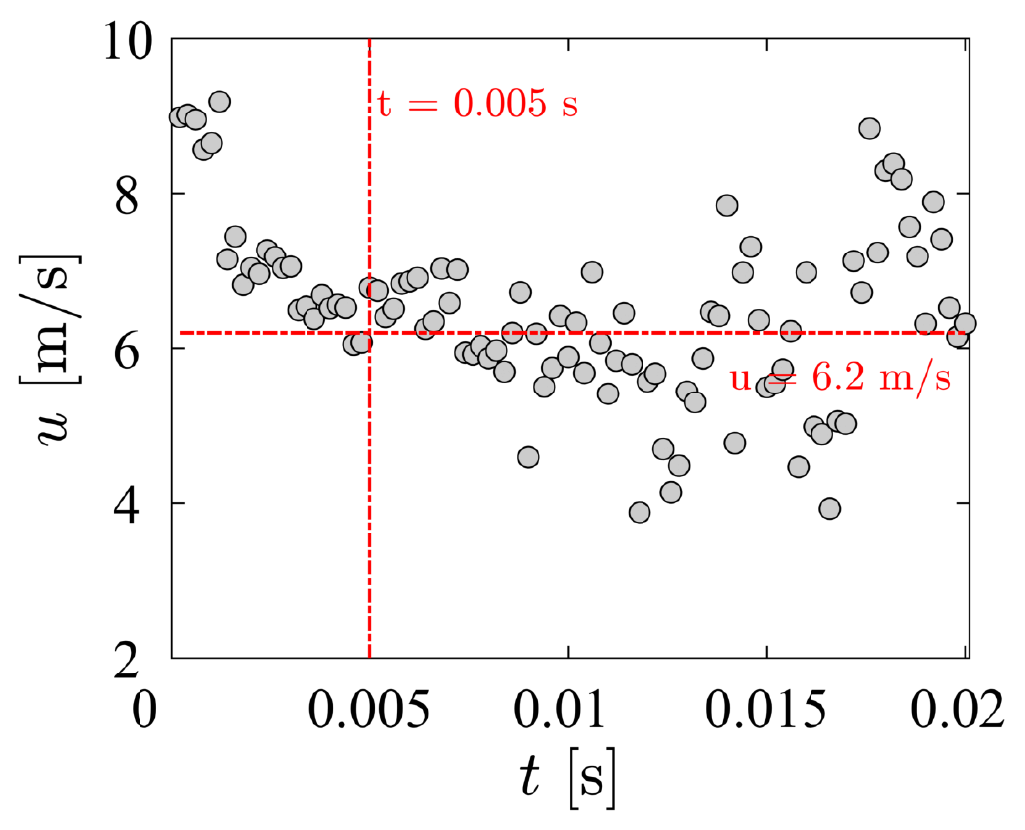}
		\quad
		\includegraphics[height=0.3\textwidth]{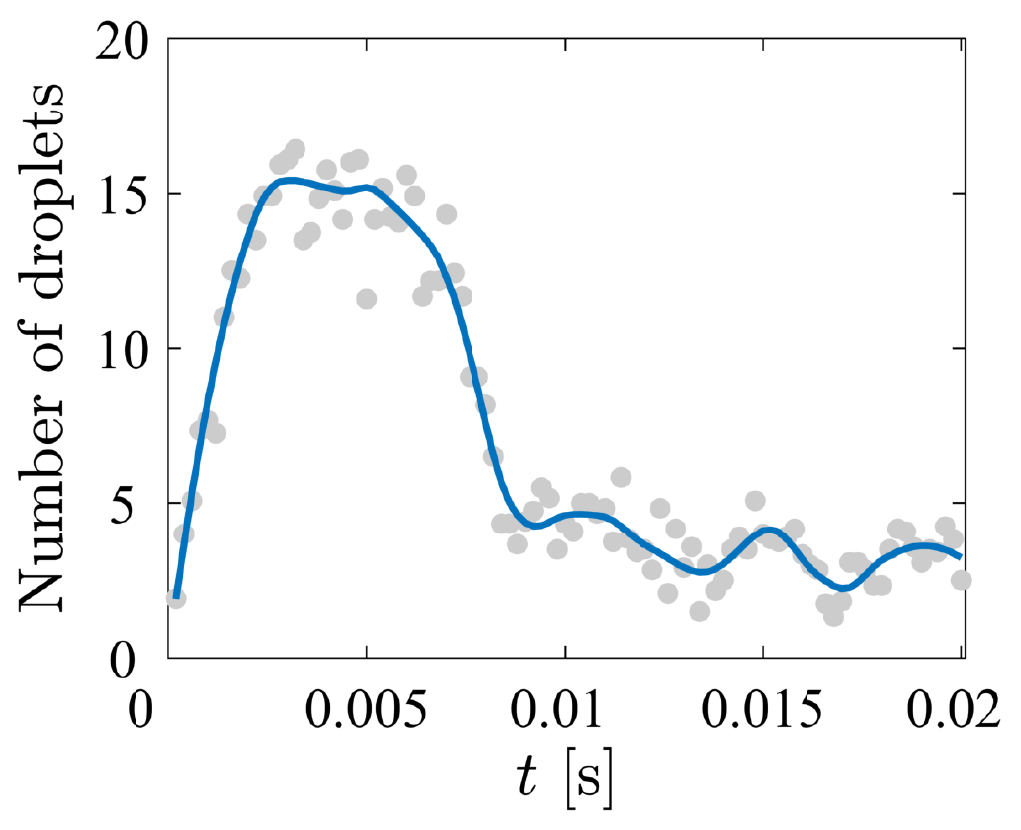}
		\put(-275,-9){(a)}
		\put(-90,-9){(b)}
\end{center}
\caption{(a) Average droplet velocity as a function of time. (b) Number of expelled droplets as a function of time in the measurement domain.}\label{fig:droplets_vel}
\end{figure*}

As introduced in Sec. \ref{subsec:PSV}, the velocity of saliva droplets is estimated using PSTV. There are 12 cases overall. The onset of coughing is visually determined as when the droplets just begin to enter the measurement domain. Because the domain is as small as approximately 1.4 cm $\times$ 2.8 cm, the velocity of the droplets is averaged over the domain and cases, and the result is presented in Fig. \ref{fig:droplets_vel} (a). At the beginning ($t$ = 0 s), the velocity of the droplets is approximately 9.0 m/s; the velocity has reduced to 6.2 m/s at $t$ = 0.005 s. After $t$ = 0.005 s, the velocity of the droplets presents large fluctuations around the mean value of 6.2 m/s. This large deviation may be caused by an insufficient number of particles, as shown in Fig. \ref{fig:droplets_vel} (b). Because the camera used cannot resolve smaller droplets, the maximum number of detected droplets in the domain is approximately 15 and decreases to approximately 3 droplets after $t$ = 0.01 s.

\begin{figure}
\begin{center}
		\includegraphics[height=0.3\textwidth]{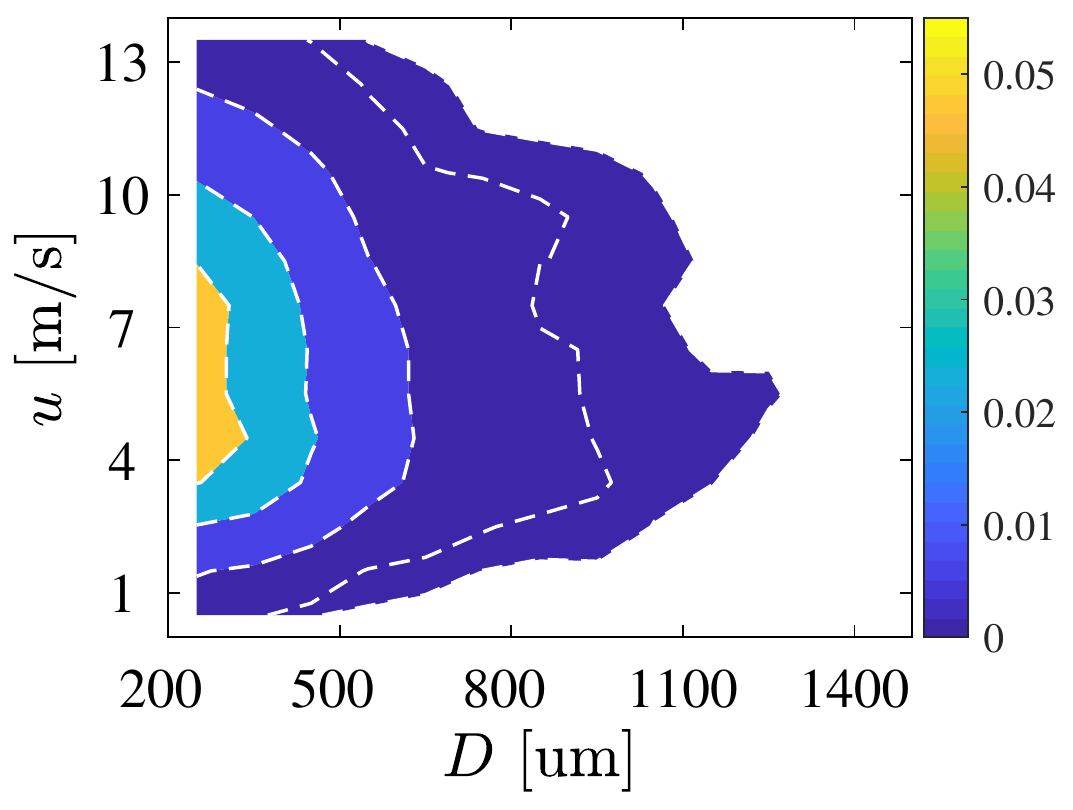}
\end{center}
\caption{Joint probability density function (PDF) of the droplet velocity and diameter.}\label{fig:droplets_jpdf}
\end{figure}

We also consider the joint probability density function (PDF) of the droplet velocity and diameter, as shown in Fig. \ref{fig:droplets_jpdf}. This joint PDF is counted over $t$ = 0 s to $t$ = 0.02 s for all cases. The droplet diameter is estimated from the equivalent area of the identified droplets. The minimum diameter is approximately 250 $\mu$m, which is 7 pixels in accordance with the image resolution of 34.6 $\mu$m/pixel. Therefore, we only measure the velocities of the large droplets. Most of the droplet velocities are distributed in the range of 4-8 m/s. The width of the velocity distribution increases as the diameter $D$ decreases because the smaller droplets are more susceptible to the velocity of the airflow. The velocity of the larger droplets is approximately 6.0 m/s.

\subsection{Physical model of the droplet movement}\label{subsec:model}

We consider a saliva droplet as a pure water sphere with diameter $D$. According to Newton's second law of motion, the evolution of the saliva droplet's velocity can be calculated as:
\begin{equation}\label{eq:vel}
	m_d\frac{d\mathbf{u}_d}{dt} = {\mathbf{F}_g}+{\mathbf{F}_a},
\end{equation}
where $m_d$ and $\mathbf{u}_d$ are the mass and velocity vectors of the droplet, respectively. The mass $m_d$ is estimated as $1/6\pi\rho_dD^3$, where $\rho_d$ is the density of the droplet. $\mathbf{F}_g$ denotes the gravity and $\mathbf{F}_a$ denotes the air drag force. In the CFD analysis of \cite{Zhu2006}, the pressure force is also considered. However, this term is ignored in Eq. \ref{eq:vel} because an accurate pressure field is difficult to obtain in our experiments. The forces $\mathbf{F}_g$ and $\mathbf{F}_a$ are given as:
\begin{equation}\label{eq:force}
	\begin{split}
		\mathbf{F}_g &= \frac{1}{6}\pi\rho_dD^3(\rho_d-\rho_a)\mathbf{g}\qquad\mathbf{g}=(0,-9.8)m/s^2, \\
		\mathbf{F}_a &= \frac{1}{2}C_a\rho_aA_d|\mathbf{u}_a-\mathbf{u}_d|(\mathbf{u}_a-\mathbf{u}_d).
	\end{split}
\end{equation}
Here, $\mathbf{g}$ is the gravitational acceleration and $A_d$ is the windward area, given as $\pi(\frac{D}{2})^2$. The parameter $\rho_a$ is the density of the air. \cite{White2006} offered the following curve-fit function for the drag coefficient $C_a$.
\begin{equation}\label{eq:Cp}
	C_a = \frac{24}{Re} + \frac{6}{1+\sqrt{Re}} + 0.4,\quad 0\leq Re \leq 2\times10^5.
\end{equation}
The Reynolds number of the droplet is computed by
\begin{equation}\label{eq:Re}
	Re = \frac{\rho_a|\mathbf{u}_a-\mathbf{u}_d|D}{\mu_a},
\end{equation}
where $\mu_a$ is the dynamic viscosity of the air. If the location of the droplet is in the region of the cough airflow, the velocity $\mathbf{u}_a$
is given as Eq. \ref{eq:uc_s}. Otherwise, the velocity $\mathbf{u}_a$ is set to zero for still ambient air. Eqs. \ref{eq:force}, \ref{eq:Cp}, \ref{eq:Re} and \ref{eq:uc_s} are substituted into Eq. \ref{eq:vel} to calculate the velocity and trajectory of the droplets.

When a respiratory droplet is expelled into the air, the physical processes of mass transfer and heat transfer are simultaneously generated at the droplet surface \cite[]{Kukkonen1989,Xie2007}. Mass reduction due to evaporation has a significant influence on the velocity and trajectory of small droplets. The rate of decrease in the diameter $D$ of a spherical drop in air due to evaporation is expressed as \cite[]{Kukkonen1989,Holterman2003}:
\begin{equation}\label{eq:evap}
	\frac{dD}{dt} = \frac{4M_LD_{\infty}P_t(1+0.276Re^{1/2}Sc^{1/3})}{RT_{\infty}}\ln\left[\frac{1-p_{sat}\left(T_w\right)/{P_t}}{1-RH\cdot p_{sat}\left(T_{\infty}\right)/{P_t}}\right].
\end{equation}
Here, $M_L$ is the molecular weight of vapor, which is given as 0.018 kg/mole; $D_\infty$ is the binary diffusion coefficient far from the droplet; $P_t$ denotes the atmospheric pressure of air (101 kPa in this work); $R$ is the universal gas constant (R = 8.3144 J mol$^{-1}$K$^{-1}$); $T_\infty$ is the ambient air temperature far from the droplet; $RH$ is the relative humidity of the ambient air; $T_w$ is the wet-bulb temperature and $p_{sat}$ denotes the saturated vapor pressure. The Schmidt number $Sc$ is a dimensionless quantity relating the momentum diffusivity (kinematic viscosity) to the mass diffusivity and is calculated by:
\begin{equation}\label{eq:Sc}
	Sc = \frac{\mu_a}{\rho_a D_\infty}.
\end{equation}
The estimation of $D_\infty$, $T_w$ and $p_{sat}$ and the validation of this model are given in detail in Sec. \ref{app:evap} of the appendix. In the present work, the temperature and relative humidity of the cough airflow are $T_{cough}$ = 34$^{\circ}$C and $RH_{cough}$ = 85$\%$, and these values for ambient air are $T_{air}$ = 23$^{\circ}$C and $RH_{air}$ = 50$\%$ \cite[]{Bourouiba2014}. The height of mouth from the ground is set to 1.8 m. The calculation is performed until the diameter is less than 5 $\mu$m. The temperature variation of the droplet is neglected in this work.

\begin{figure*}
\begin{center}
		\includegraphics[height=0.35\textwidth]{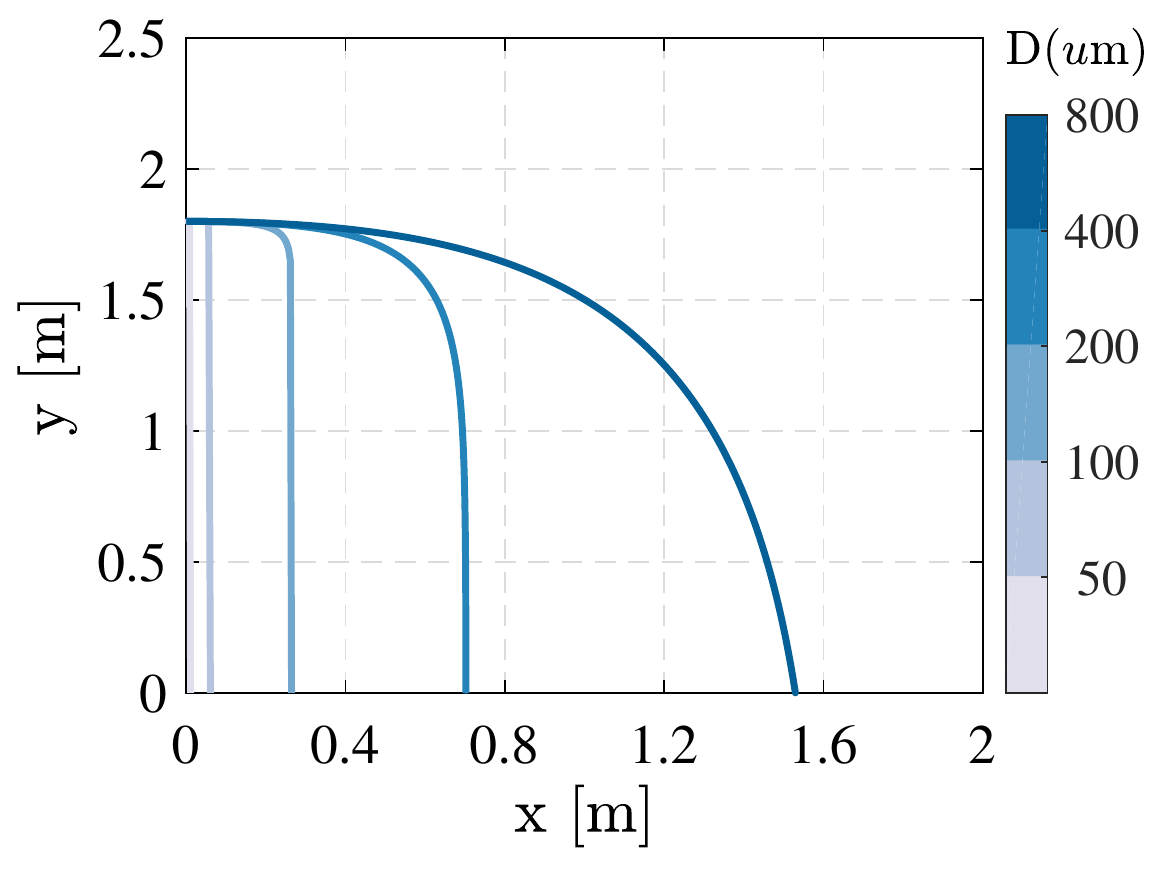}
		\quad
		\includegraphics[height=0.35\textwidth]{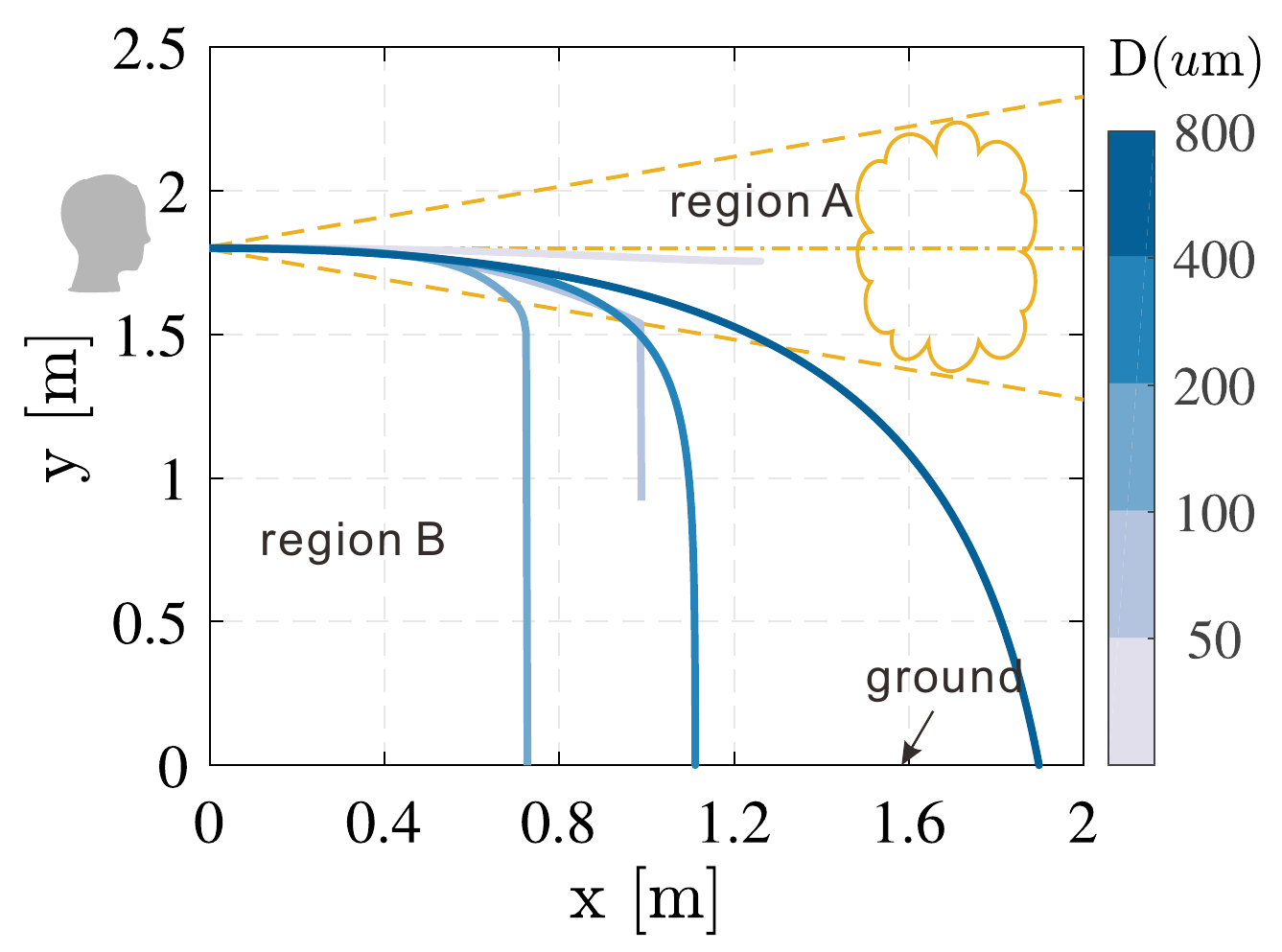}
		\put(-355,-9){(a)}
		\put(-120,-9){(b)}
\end{center}
\caption{Trajectories of the droplets at five different diameters of $D$ = 30 $\mu$m, 80 $\mu$m, 200 $\mu$m, 400 $\mu$m, 800 $\mu$m (a) without considering the cough airflow and evaporation effect, (b) considering the cough airflow and the evaporation effect. }\label{fig:model_with_cough}
\end{figure*}

Figure \ref{fig:model_with_cough} displays the trajectories of droplets at the five different diameters of $D$ = 30 $\mu$m, 80 $\mu$m, 200 $\mu$m, 400 $\mu$m, 800 $\mu$m. The simulated results considering the cough airflow and evaporation are shown in Fig. \ref{fig:model_with_cough} (b). For comparison, only gravity and the drag force are considered in Fig. \ref{fig:model_with_cough} (a). The initial horizontal velocity of the droplets is 6 m/s. It can be seen from Fig. \ref{fig:model_with_cough} (a) that the horizontal velocity rapidly decreases to zero for the small droplet due to its relatively high viscosity. All the droplets fall to the ground, and the distance from the cougher at which this occurs $x$ increases with increasing diameter. However, considering the real physical process, there are three different types of trajectories, as shown in Fig. \ref{fig:model_with_cough} (b). First, the trajectory for a droplet of $D$ = 30 $\mu$m stops in the region of the cough airflow, which is denoted region A. The diameter of this droplet deceases to 5 $\mu$m due to evaporation. Second, the trajectory for a droplet of $D$ = 80 $\mu$m stops in the region of ambient air, which is denoted region B. Third, droplets with larger diameters of $D$ = 200 $\mu$m, 400 $\mu$m and 800 $\mu$m fall to the ground before evaporating to droplet nuclei. Moreover, the droplets in Fig. \ref{fig:model_with_cough} (b) travel longer distances than those in Fig. \ref{fig:model_with_cough} (a). Taking $D$ = 200 $\mu$m as an example, the cough airflow can increase the distance from 0.25 m to 0.7 m.

\begin{figure*}
\begin{center}
		\includegraphics[height=0.35\textwidth]{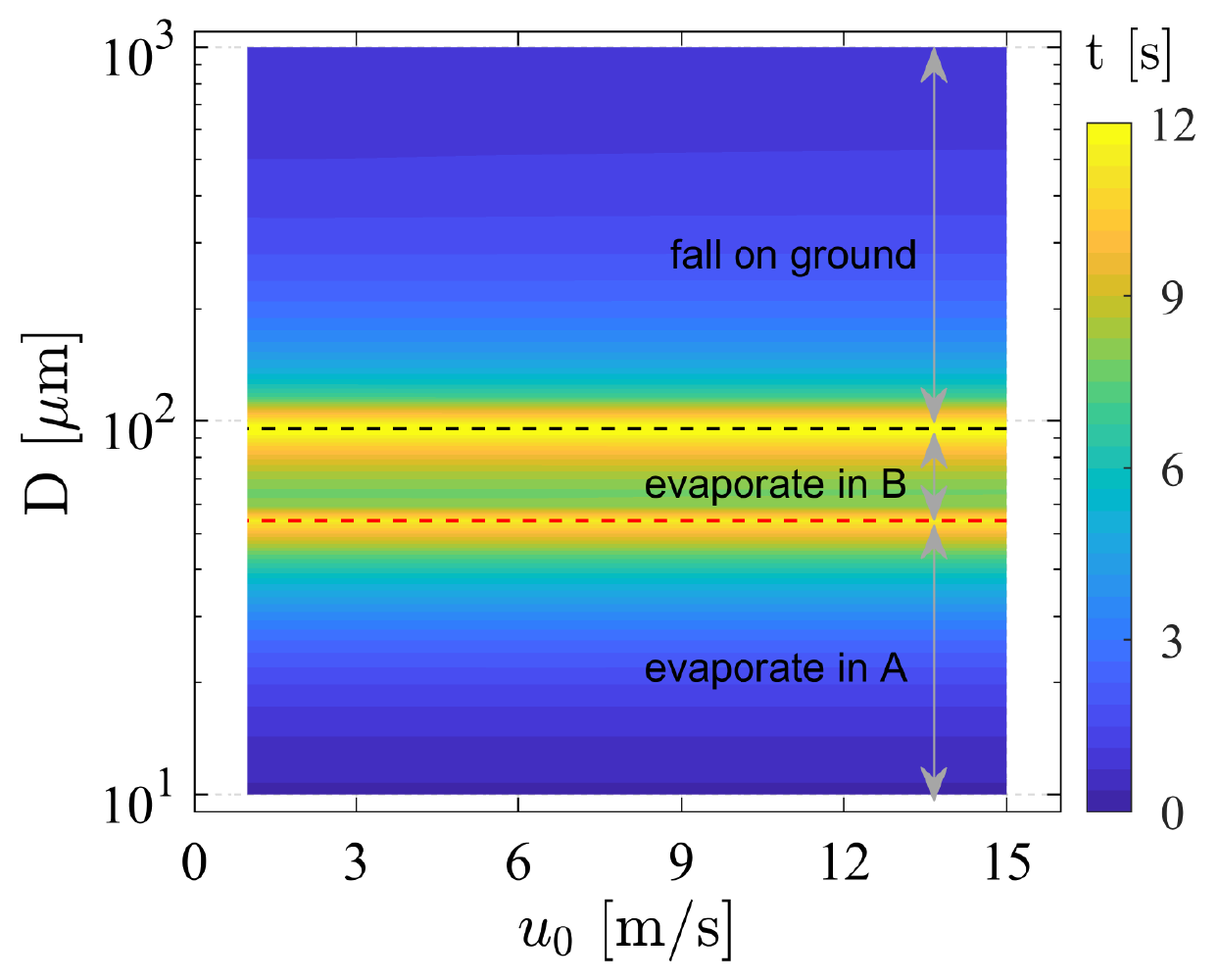}
		\includegraphics[height=0.35\textwidth]{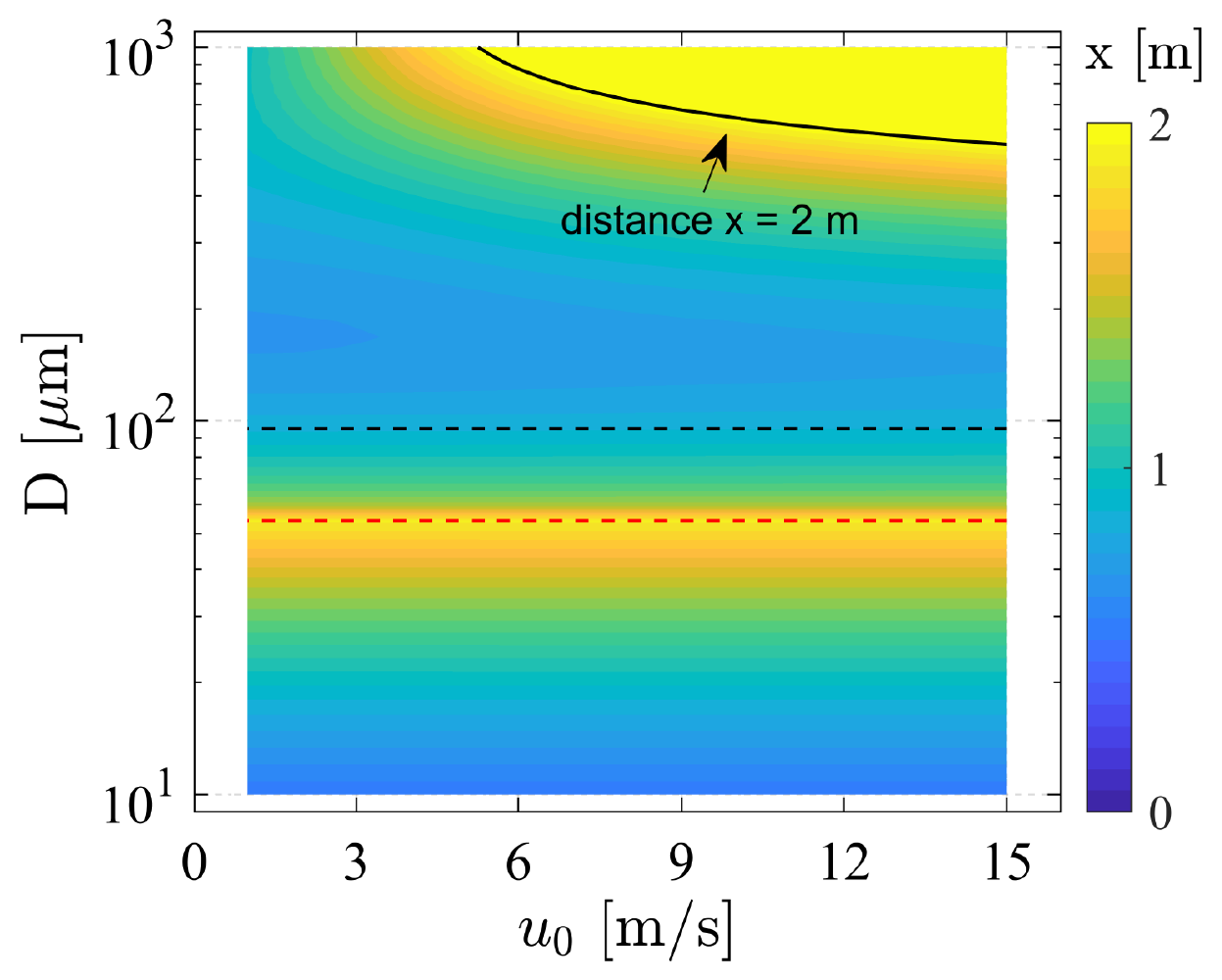}
		\put(-315,-9){(a)}
		\put(-105,-9){(b)}
\end{center}
\caption{(a) Evaporation time and falling time as a function of the droplet diameter and initial velocity. (b) Distance traveled from the cougher (to evaporate to a small droplet with $D$ = 5 $\mu$m or to fall to the ground) as a function of the droplet diameter and initial velocity. The distinction between the droplets that fall to the ground and those that evaporate to droplet nuclei is represented by a black dashed line, and the distinction between the droplets that evaporate in region A and those that evaporate in region B is denoted by a red dashed line.}\label{fig:model_cough_time}
\end{figure*}

The evaporation time and falling time as a function of the droplet diameter and initial velocity are presented in Fig. \ref{fig:model_cough_time} (a). There are two dashed lines in this figure: the black line represents the dividing line between droplets that fall to the ground and those that evaporate to droplet nuclei, and the red line represents the dividing line between the droplets that evaporate in region A and those that evaporate in region B. With our parameter settings, droplets larger than the critical droplet size $D \approx 100$ $\mu$m are deposited on the ground before evaporating to a small droplet of $D$ = 5 $\mu$m. This critical droplet size is consistent with the results given by \cite{Wells1955} and \cite{Xie2007}. The second critical size indicated by the red dashed line is approximately 50 $\mu$m, where droplets smaller than this size evaporate to droplet nuclei in region A (cough airflow). The droplets with diameters of 50 $\sim$ 100 $\mu$m evaporate in region B (ambient air). These droplet nuclei are suspended in the air and travel with the movement of the air. In particular, the droplets in region A are easier for people of the same height to inhale; therefore, these droplets pose a higher probability of infection. Moreover, the initial velocity of the droplets $u_0$ has a negligible influence on the time.

The horizontal distance covered as a droplet evaporates to a small droplet or falls to the ground is displayed in Fig. \ref{fig:model_cough_time} (b). The contour line of $x$ = 2 m is indicated by a black arrow. The contour map above the black dashed line represents the distance at which the droplet falls to the ground, and the contour map lower the back dashed line represents the distance traveled before evaporation. The practical distance of the droplets with $D$ $\leq$ 100 $\mu$m is hard to predict because of the long-range airborne transmission. In addition to the small droplets, the droplets with $D$ $\geq$ 500 $\mu$m and $u_0 \geq$  5 m/s can travel further than 2 m, as shown in Fig. \ref{fig:model_cough_time} (b). A social distance of 2 m cannot completely eliminate the possibility of infection; thus, wearing a mask is a simple and direct way to stop the spread of droplets.

\begin{figure}
\begin{center}
		\includegraphics[height=0.35\textwidth]{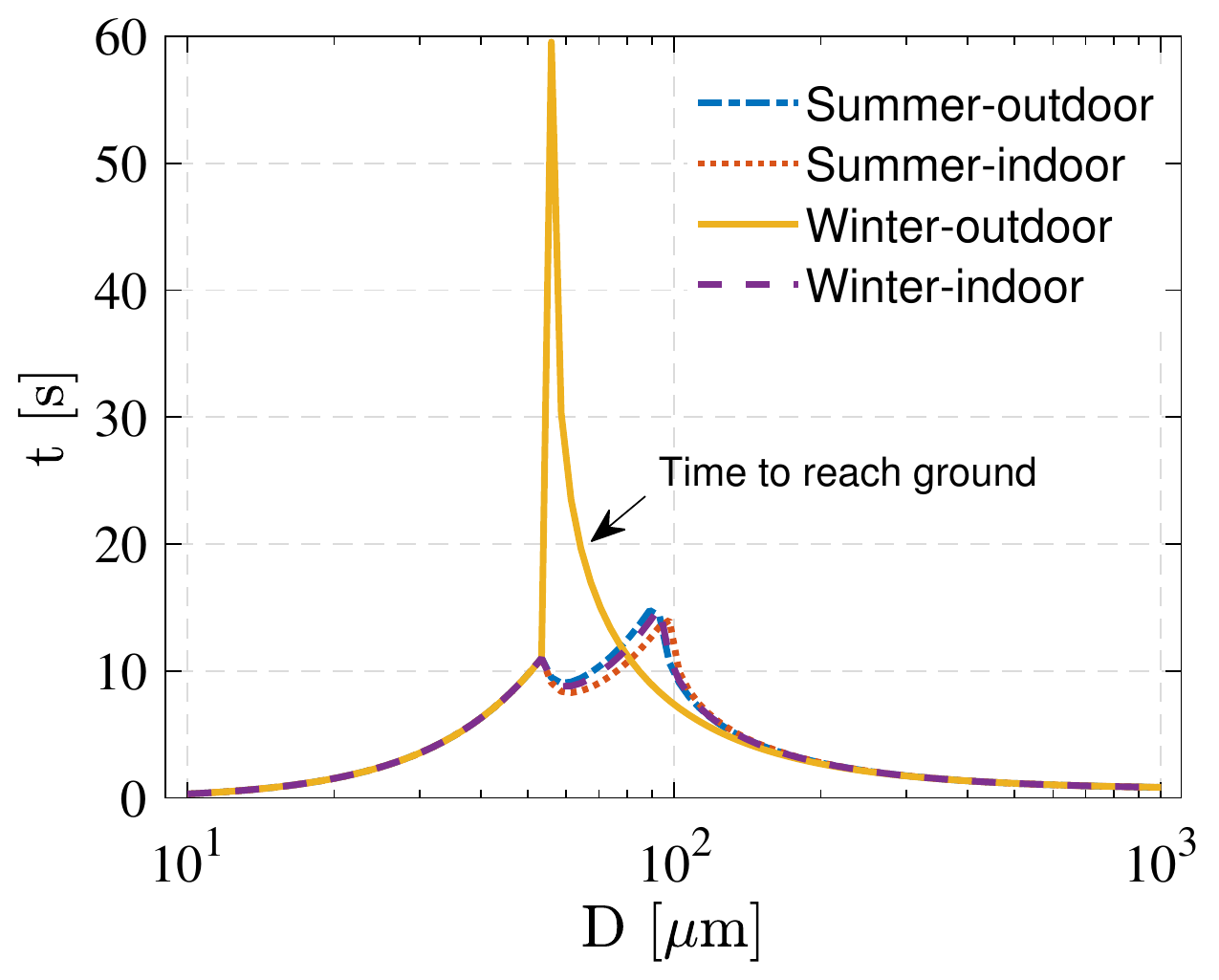}
\end{center}
\caption{Evaporation-falling times for different weather conditions.}\label{fig:model_time_season}
\end{figure}

With this model, we can investigate the effects of weather conditions on the evaporation of coughed droplets. Measurements including the seasonal variations in the temperature and relative humidity of indoor and outdoor conditions are given by \cite{Frankel2012}. The median values of the temperature and relative humidity measured during summer and winter are listed in Tab. \ref{tab:season}. Substituting these parameters into our model, we can obtain the evaporation-falling time curves of the droplets under different weather conditions. The results for an initial droplet velocity of 6 m/s are shown in Fig. \ref{fig:model_time_season}. The curves of the summer-outdoor, summer-indoor and winter-indoor areas almost overlap each other due to their similar temperatures and relative humidities. However, for winter-outdoor conditions, droplets with initial diameters larger than 50 $\mu$m evaporate slowly and deposit on the ground before drying. Additionally, the settling time of droplets with diameters of 50 $\sim$ 100 $\mu$m is much longer than those of the other three cases. If we consider the effect of condensation in winter, more large droplets are slowly settling to the ground in these conditions. This finding indicates that low temperature and high relative humidity may significantly increase the possibility of large droplet transmission and contact transmission, which may be a potential reason for a second pandemic wave in the autumn and winter seasons \cite[]{Dbouk2020_2}.

\begin{table}
	\caption{Temperatures and relative humidities measured during summer and winter \cite[]{Frankel2012}.}\label{tab:season} 
	\begin{ruledtabular}
		\begin{tabular}{ccc}
			& summer & winter\\
			\hline
			outdoor temp ($^\circ$C) & 22.4 & 0.75\\
			outdoor RH (\%) & 62.0 & 97.0 \\
			indoor temp ($^\circ$C) & 24.2 & 19.3\\
			indoor RH (\%) & 57.3 & 55.7 \\
		\end{tabular}
	\end{ruledtabular}
\end{table}

\section{Conclusions}\label{sec:con}
In the present work, the motion of saliva droplets produced by coughing was investigated through experiments and theoretical analysis.

Three different types of experiments, including flow visualization, PIV and PSTV, were performed to experimentally investigate the transport characteristics of coughing. For the flow visualization and PIV, image sequences of cigarette smoke expelled by healthy adult male volunteers were recorded using a high-speed camera. For the PSTV, large droplets were recorded using a shadow imaging technique, and the velocity and diameter of these droplets were extracted from the images by tracking the motion of the droplets. Our key findings from the experiments are as follows. First, the visualization shows that the cough jet interacts with the ambient air via flow entrainment, which leads to increasing size and decreasing velocity with increasing distance from the cougher \cite[]{Bourouiba2014}. Further quantitative analysis indicates that the convection velocity of the cough airflow presents the relationship $t^{-0.7}$ with time; hence, the distance from the cougher increases as $t^{0.3}$ in the range of our measurement domain. The dependence of the distance $s$ on time $t$ is different from the relationship $s\sim t^{0.5}$ that was obtained by considering a cough as a continuous jet by \cite{Bourouiba2014}. The width of the cough airflow evolves linearly as the distance increases. Second, the normalized mean velocity profiles agrees well with that of a synthetic jet in the region $y/b_{1/2} \leq 0$ and is higher than that of a synthetic jet in the region $y/b_{1/2}$ > 0, mainly due to the buoyancy driven by the relatively high temperature of a cough. The maximum airflow can reach 15.0 m/s. Third, the measured minimum diameter is approximately 250 $\mu$m due to the limitations of our experimental hardware. The mean velocity of the droplets is approximately 6.0 m/s, and smaller droplets are more susceptible to cough airflow.

With these experimental results, a physical model considering the evaporation effect was built to predict the movement of droplets under different environmental conditions (temperature and relative humidity). Our findings indicate that there are two critical sizes for the expelled droplets. The first critical droplet size, where droplets larger than this size fall to the ground or otherwise evaporate to droplet nuclei, is approximately 100 $\mu$m. The second critical droplet size, which distinguishes droplet evaporation in the region of cough airflow from that in the region of ambient air, is approximately 50 $\mu$m. Droplets smaller than 50 $\mu$m are easier for people of the same height to inhale, resulting in a higher probability of infection. Additionally, both the small droplets (initial diameter $D \leq$ 100 $\mu$m) evaporating to droplet nuclei and the large droplets with $D \geq$ 500 $\mu$m and initial velocity $u_0 \geq$ 5 m/s can travel more than the social distance of 2 m. With this model, we also investigated the effects of weather conditions on the evaporation of cough droplets. Winter conditions, with low temperature and high relative humidity, can result in more droplets settling to the ground, which may be a possible driver of a second pandemic wave in the autumn and winter seasons.

Our study visualizes and analyzes the cough process using techniques of experimental fluid mechanics, and the theoretical results reinforce the importance of maintaining social distance and wearing masks to stem the spread of viruses. We should take precautions against a second COVID-19 wave due to the winter conditions of low temperature and high relative humidity. A potential future study is to build a relationship between pandemic evolution and the motion of respiratory droplets to accurately predict the infection rate and range for the population.

\begin{acknowledgments}
This work was supported by the NSFC Basic Science Center Program for `Multiscale Problems in Nonlinear Mechanics' (Grant No. 11988102) and by the National Natural Science Foundation of China (Grant No. 11702302). The authors would also like to acknowledge the support from the Strategic Priority Research Program (Grant No. XDB22040104) and the Key Research Program of Frontier Sciences of the Chinese Academy of Sciences (Grant No. QYZDJ-SSW-SYS002). The authors wish to give special thanks to Xiaodong Chen for helpful discussions on the evaporation model used in the manuscript.
\end{acknowledgments}

\section*{DATA AVAILABILITY}
The data that support the findings of this study are available from the corresponding authors upon reasonable request.

\appendix
\section{The evaporation model} \label{app:evap}
\begin{figure}
\begin{center}
		\includegraphics[height=0.4\textwidth]{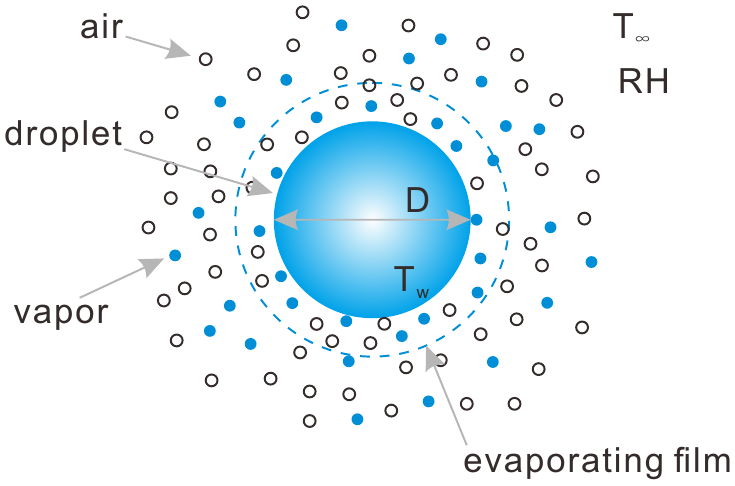}
\end{center}
\caption{Schematic diagram of droplet evaporation in air. The open and solid circles represent air and vapor, respectively. }\label{fig:sche_evap}
\end{figure}
A schematic representation of droplet evaporation is shown in Fig. \ref{fig:sche_evap}. In this figure, the open and solid circles represent air and vapor, respectively. The ambient air temperature and relative humidity are denoted by $T_\infty$ and $RH$, respectively. The molecules in the droplet are bound to their neighbors by intermolecular forces; however, the liquid molecules at the interface are more weakly bounded to their neighbors than those inside the liquid due to having fewer adjacent molecules \cite[]{Holterman2003}. Therefore, some molecules with relatively high kinetic energy escape from the droplet into the air, and a very thin evaporating film with a saturated vapor is generated at the interface between the droplet and air. Because escaped molecules can absorb heat, the droplet's temperature $T_w$ is lower than the temperature of the ambient air $T_\infty$. In this work, we ignore the variation in the temperature of the droplet and assume that the final droplet temperature is the wet-bulb temperature, which can be described by a linear relation with $T_\infty$ and quadratic relation with relative humidity $RH$ as \cite[]{Holterman2003}:
\begin{equation}\label{eq:tw}
	\begin{split}
		T_w &= T_\infty-\left[\left(a_0+a_1T_\infty\right)+\left(b_0+b_1T_\infty\right)RH+\left(c_0+c_1T_\infty\right)RH^2\right], \\
		a_0 &= 5.1055 \qquad a_1 = 0.4295,	\\
		b_0 &= -0.04703 \qquad b_1 = -0.005951, \\
		c_0 &= -4.005\times10^{-5} \qquad c_1 = 1.660\times10^{-5}.
	\end{split}
\end{equation}
At 100$\%$ relative humidity, the wet-bulb temperature $T_w$ is equal to the air temperature $T_\infty$ (dry-bulb temperature).

If the air is saturated with the water vapor, the partial vapor pressure is defined as the saturated vapor pressure $p_{sat}$. $p_{sat}$ is strongly dependent on the temperature, and an empirical equation is given as \cite[]{Holterman2003}:
\begin{equation}
	p_{sat} = 610.7\times10^{7.5T/(T+237.3)}.
\end{equation}
The above equation is valid for temperatures of 0$^\circ$C up to approximately 100$^\circ$C.

The dependence of the diffusion coefficient $D_\infty$ of water in air on the temperature is given by \cite{Holterman2003} as:
\begin{equation}
	D_\infty = 21.2\times10^{-6}(1+0.0071T_\infty).
\end{equation}

\begin{figure*}
\begin{center}
		\includegraphics[height=0.35\textwidth]{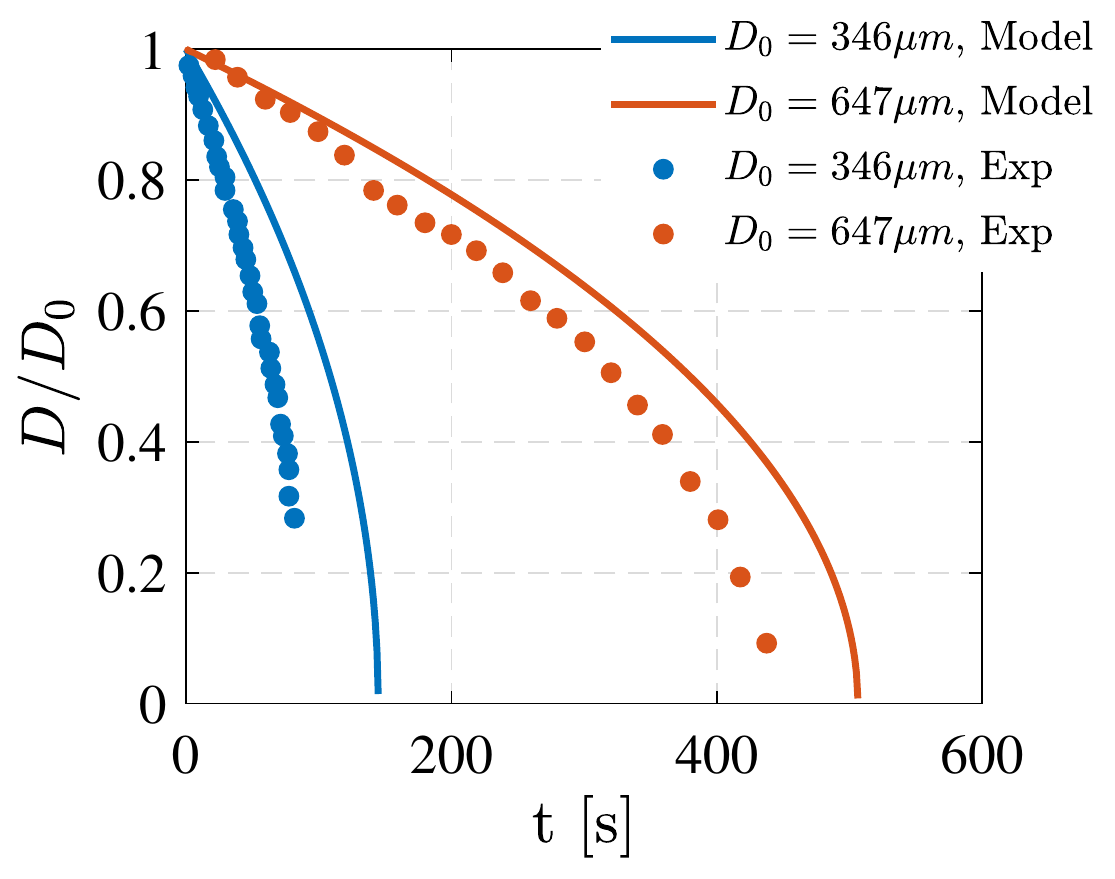}
\quad
		\includegraphics[height=0.35\textwidth]{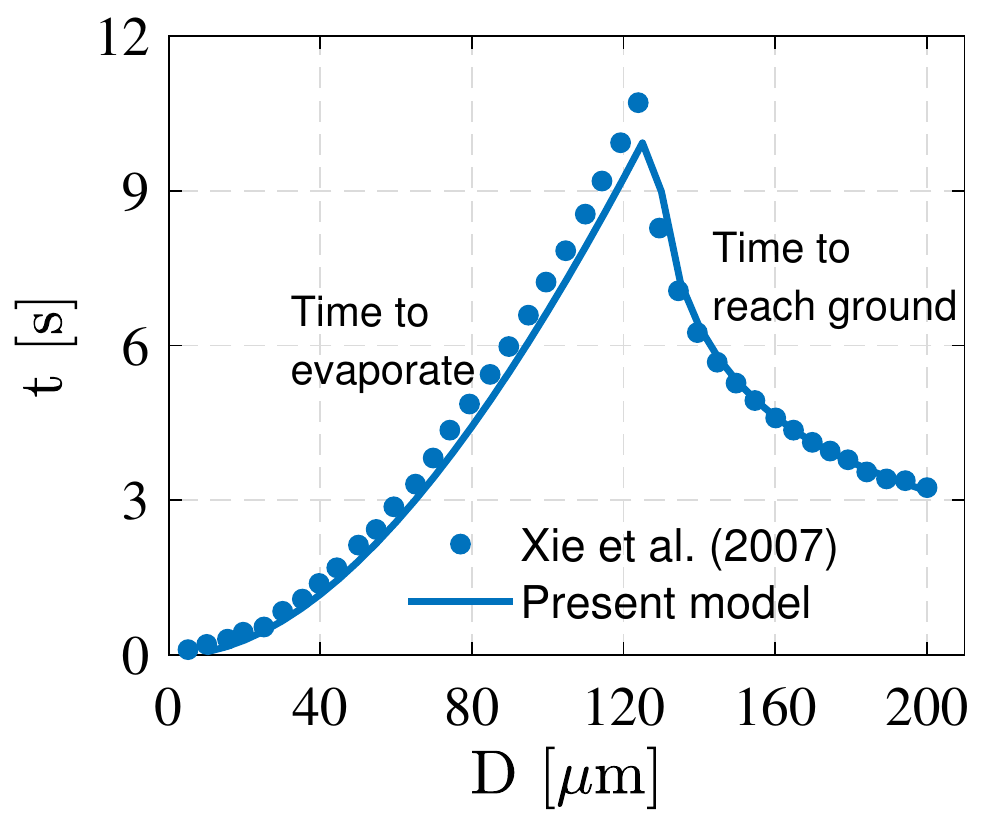}
\put(-315,-9){(a)}
\put(-90,-9){(b)}
\end{center}
\caption{(a) Comparison between the present model and the experimental results given by \cite{Chaudhuri2020}. The diameter evolution of the motionless droplet is conducted at $T_\infty$ = 30$^\circ$C and $RH$ = 50$\%$. $D_0$ is the initial diameter of the droplet. (b) Comparison of the evaporation-falling time between the present model and the result extracted from the paper by \cite{Xie2007}. The ambient temperature $T_\infty$ is set to 18$^\circ$C, and the relative humidity $RH$ is set to $0\%$. The droplet is released at a height of 2 m. }\label{fig:model_validation}
\end{figure*}

The validity of the present model is first examined by comparing the diameter evolution of motionless droplets with the experimental results given by \cite{Chaudhuri2020} (Fig. \ref{fig:model_validation} (a)). A motionless droplet is allowed to evaporate under ambient conditions at $T_\infty$ = 30$^\circ$C and $RH$ = 50$\%$. The diameter variations as a function of time for two different initial diameters $D_0$ are shown in Fig. \ref{fig:model_validation} (a). The solid curves estimated by the present model qualitatively agree with the experimental results. We also compare the evaporation-falling time of the present model with the result extracted from the paper by \cite{Xie2007}. The droplet is released at a height of 2 m, and the ambient temperature $T_\infty$ and relative humidity $RH$ are 18$^\circ$C and $0\%$, respectively. As shown in Fig. \ref{fig:model_validation} (b), the curve estimated by the present model agrees well with the result of \cite{Xie2007}. Droplets smaller than 125 $\mu$m totally evaporate before reaching the ground. With these comparisons, we can conclude that the model used in this work can describe the evaporation and motion of a droplet.

\nocite{*}
\bibliography{References}% Produces the bibliography via BibTeX.

%merlin.mbs aipauth4-1.bst 2010-07-25 4.21a (PWD, AO, DPC) hacked
%Control: author (9) reversed initials
\begin{thebibliography}{43}%
\makeatletter
\providecommand \@ifxundefined [1]{%
 \@ifx{#1\undefined}
}%
\providecommand \@ifnum [1]{%
 \ifnum #1\expandafter \@firstoftwo
 \else \expandafter \@secondoftwo
 \fi
}%
\providecommand \@ifx [1]{%
 \ifx #1\expandafter \@firstoftwo
 \else \expandafter \@secondoftwo
 \fi
}%
\providecommand \natexlab [1]{#1}%
\providecommand \enquote  [1]{``#1''}%
\providecommand \bibnamefont  [1]{#1}%
\providecommand \bibfnamefont [1]{#1}%
\providecommand \citenamefont [1]{#1}%
\providecommand \href@noop [0]{\@secondoftwo}%
\providecommand \href [0]{\begingroup \@sanitize@url \@href}%
\providecommand \@href[1]{\@@startlink{#1}\@@href}%
\providecommand \@@href[1]{\endgroup#1\@@endlink}%
\providecommand \@sanitize@url [0]{\catcode `\\12\catcode `\$12\catcode
  `\&12\catcode `\#12\catcode `\^12\catcode `\_12\catcode `\%12\relax}%
\providecommand \@@startlink[1]{}%
\providecommand \@@endlink[0]{}%
\providecommand \url  [0]{\begingroup\@sanitize@url \@url }%
\providecommand \@url [1]{\endgroup\@href {#1}{\urlprefix }}%
\providecommand \urlprefix  [0]{URL }%
\providecommand \Eprint [0]{\href }%
\providecommand \doibase [0]{http://dx.doi.org/}%
\providecommand \selectlanguage [0]{\@gobble}%
\providecommand \bibinfo  [0]{\@secondoftwo}%
\providecommand \bibfield  [0]{\@secondoftwo}%
\providecommand \translation [1]{[#1]}%
\providecommand \BibitemOpen [0]{}%
\providecommand \bibitemStop [0]{}%
\providecommand \bibitemNoStop [0]{.\EOS\space}%
\providecommand \EOS [0]{\spacefactor3000\relax}%
\providecommand \BibitemShut  [1]{\csname bibitem#1\endcsname}%
\let\auto@bib@innerbib\@empty
%</preamble>
\bibitem [{\citenamefont {Asadi}\ \emph {et~al.}(2020)\citenamefont {Asadi},
  \citenamefont {Bouvier}, \citenamefont {Wexler},\ and\ \citenamefont
  {Ristenpart}}]{Asadi2020}%
  \BibitemOpen
  \bibfield  {author} {\bibinfo {author} {\bibnamefont {Asadi}, \bibfnamefont
  {S.}}, \bibinfo {author} {\bibnamefont {Bouvier}, \bibfnamefont {N.}},
  \bibinfo {author} {\bibnamefont {Wexler}, \bibfnamefont {A.~S.}}, \ and\
  \bibinfo {author} {\bibnamefont {Ristenpart}, \bibfnamefont {W.~D.}},\
  }\bibfield  {title} {\enquote {\bibinfo {title} {{The coronavirus pandemic
  and aerosols: Does COVID-19 transmit via expiratory particles?}}}\ }\href
  {\doibase 10.1080/02786826.2020.1749229} {\bibfield  {journal} {\bibinfo
  {journal} {Aerosol Science and Technology}\ ,\ \bibinfo {pages} {4}}
  (\bibinfo {year} {2020})}\BibitemShut {NoStop}%
\bibitem [{\citenamefont {Bourouiba}, \citenamefont {Dehandschoewercker},\ and\
  \citenamefont {Bush}(2014)}]{Bourouiba2014}%
  \BibitemOpen
  \bibfield  {author} {\bibinfo {author} {\bibnamefont {Bourouiba},
  \bibfnamefont {L.}}, \bibinfo {author} {\bibnamefont {Dehandschoewercker},
  \bibfnamefont {E.}}, \ and\ \bibinfo {author} {\bibnamefont {Bush},
  \bibfnamefont {J.~W.~M.}},\ }\bibfield  {title} {\enquote {\bibinfo {title}
  {Violent expiratory events: on coughing and sneezing},}\ }\href {\doibase
  10.1017/jfm.2014.88} {\bibfield  {journal} {\bibinfo  {journal} {Journal of
  Fluid Mechanics}\ }\textbf {\bibinfo {volume} {745}},\ \bibinfo {pages}
  {537--563} (\bibinfo {year} {2014})}\BibitemShut {NoStop}%
\bibitem [{\citenamefont {Castanet}\ \emph {et~al.}(2013)\citenamefont
  {Castanet}, \citenamefont {Dunand}, \citenamefont {Caballina},\ and\
  \citenamefont {Lemoine}}]{Castanet2013}%
  \BibitemOpen
  \bibfield  {author} {\bibinfo {author} {\bibnamefont {Castanet},
  \bibfnamefont {G.}}, \bibinfo {author} {\bibnamefont {Dunand}, \bibfnamefont
  {P.}}, \bibinfo {author} {\bibnamefont {Caballina}, \bibfnamefont {O.}}, \
  and\ \bibinfo {author} {\bibnamefont {Lemoine}, \bibfnamefont {F.}},\
  }\bibfield  {title} {\enquote {\bibinfo {title} {High-speed shadow imagery to
  characterize the size and velocity of the secondary droplets produced by drop
  impacts onto a heated surface},}\ }\href {\doibase 10.1007/s00348-013-1489-3}
  {\bibfield  {journal} {\bibinfo  {journal} {Experiments in Fluids}\ }\textbf
  {\bibinfo {volume} {54}},\ \bibinfo {pages} {1489} (\bibinfo {year}
  {2013})}\BibitemShut {NoStop}%
\bibitem [{\citenamefont {Chaudhuri}\ \emph {et~al.}(2020)\citenamefont
  {Chaudhuri}, \citenamefont {Basu}, \citenamefont {Kabi}, \citenamefont
  {Unni},\ and\ \citenamefont {Saha}}]{Chaudhuri2020}%
  \BibitemOpen
  \bibfield  {author} {\bibinfo {author} {\bibnamefont {Chaudhuri},
  \bibfnamefont {S.}}, \bibinfo {author} {\bibnamefont {Basu}, \bibfnamefont
  {S.}}, \bibinfo {author} {\bibnamefont {Kabi}, \bibfnamefont {P.}}, \bibinfo
  {author} {\bibnamefont {Unni}, \bibfnamefont {V.~R.}}, \ and\ \bibinfo
  {author} {\bibnamefont {Saha}, \bibfnamefont {A.}},\ }\bibfield  {title}
  {\enquote {\bibinfo {title} {{Modeling the role of respiratory droplets in
  COVID-19 type pandemics}},}\ }\href {\doibase 10.1063/5.0015984} {\bibfield
  {journal} {\bibinfo  {journal} {Physics of Fluids}\ }\textbf {\bibinfo
  {volume} {32}},\ \bibinfo {pages} {063309} (\bibinfo {year}
  {2020})}\BibitemShut {NoStop}%
\bibitem [{\citenamefont {David}\ \emph {et~al.}(2012)\citenamefont {David},
  \citenamefont {Benny}, \citenamefont {Leo}, \citenamefont {Susanna},
  \citenamefont {Nelson}, \citenamefont {Tony},\ and\ \citenamefont
  {Matthew}}]{David2012}%
  \BibitemOpen
  \bibfield  {author} {\bibinfo {author} {\bibnamefont {David}, \bibfnamefont
  {S.~H.}}, \bibinfo {author} {\bibnamefont {Benny}, \bibfnamefont {K.~C.}},
  \bibinfo {author} {\bibnamefont {Leo}, \bibfnamefont {C.}}, \bibinfo {author}
  {\bibnamefont {Susanna}, \bibfnamefont {S.~N.}}, \bibinfo {author}
  {\bibnamefont {Nelson}, \bibfnamefont {L.}}, \bibinfo {author} {\bibnamefont
  {Tony}, \bibfnamefont {G.}}, \ and\ \bibinfo {author} {\bibnamefont
  {Matthew}, \bibfnamefont {T.~V.~C.}},\ }\bibfield  {title} {\enquote
  {\bibinfo {title} {{Exhaled Air Dispersion during Coughing with and without
  Wearing a Surgical or N95 Mask}},}\ }\href {\doibase
  10.1371/journal.pone.0050845} {\bibfield  {journal} {\bibinfo  {journal}
  {PLoS ONE}\ }\textbf {\bibinfo {volume} {7}},\ \bibinfo {pages} {1--7}
  (\bibinfo {year} {2012})}\BibitemShut {NoStop}%
\bibitem [{\citenamefont {Dbouk}\ and\ \citenamefont
  {Drikakis}(2020{\natexlab{a}})}]{Dbouk2020}%
  \BibitemOpen
  \bibfield  {author} {\bibinfo {author} {\bibnamefont {Dbouk}, \bibfnamefont
  {T.}}\ and\ \bibinfo {author} {\bibnamefont {Drikakis}, \bibfnamefont {D.}},\
  }\bibfield  {title} {\enquote {\bibinfo {title} {On coughing and airborne
  droplet transmission to humans},}\ }\href {\doibase 10.1063/5.0011960}
  {\bibfield  {journal} {\bibinfo  {journal} {Physics of Fluids}\ }\textbf
  {\bibinfo {volume} {32}},\ \bibinfo {pages} {10} (\bibinfo {year}
  {2020}{\natexlab{a}})}\BibitemShut {NoStop}%
\bibitem [{\citenamefont {Dbouk}\ and\ \citenamefont
  {Drikakis}(2020{\natexlab{b}})}]{Dbouk2020-1}%
  \BibitemOpen
  \bibfield  {author} {\bibinfo {author} {\bibnamefont {Dbouk}, \bibfnamefont
  {T.}}\ and\ \bibinfo {author} {\bibnamefont {Drikakis}, \bibfnamefont {D.}},\
  }\bibfield  {title} {\enquote {\bibinfo {title} {On respiratory droplets and
  face masks},}\ }\href {\doibase 10.1063/5.0015044} {\bibfield  {journal}
  {\bibinfo  {journal} {Physics of Fluids}\ }\textbf {\bibinfo {volume} {32}},\
  \bibinfo {pages} {063303} (\bibinfo {year} {2020}{\natexlab{b}})}\BibitemShut
  {NoStop}%
\bibitem [{\citenamefont {Dbouk}\ and\ \citenamefont
  {Drikakis}(2020{\natexlab{c}})}]{Dbouk2020_2}%
  \BibitemOpen
  \bibfield  {author} {\bibinfo {author} {\bibnamefont {Dbouk}, \bibfnamefont
  {T.}}\ and\ \bibinfo {author} {\bibnamefont {Drikakis}, \bibfnamefont {D.}},\
  }\bibfield  {title} {\enquote {\bibinfo {title} {Weather impact on airborne
  coronavirus survival},}\ }\href {\doibase 10.1063/5.0024272} {\bibfield
  {journal} {\bibinfo  {journal} {Physics of Fluids}\ }\textbf {\bibinfo
  {volume} {32}},\ \bibinfo {pages} {093312} (\bibinfo {year}
  {2020}{\natexlab{c}})}\BibitemShut {NoStop}%
\bibitem [{\citenamefont {Dudalski}\ \emph {et~al.}(2020)\citenamefont
  {Dudalski}, \citenamefont {Mohamed}, \citenamefont {Mubareka}, \citenamefont
  {Bi}, \citenamefont {Zhang},\ and\ \citenamefont {Savory}}]{Dudalski2020}%
  \BibitemOpen
  \bibfield  {author} {\bibinfo {author} {\bibnamefont {Dudalski},
  \bibfnamefont {N.}}, \bibinfo {author} {\bibnamefont {Mohamed}, \bibfnamefont
  {A.}}, \bibinfo {author} {\bibnamefont {Mubareka}, \bibfnamefont {S.}},
  \bibinfo {author} {\bibnamefont {Bi}, \bibfnamefont {R.}}, \bibinfo {author}
  {\bibnamefont {Zhang}, \bibfnamefont {C.}}, \ and\ \bibinfo {author}
  {\bibnamefont {Savory}, \bibfnamefont {E.}},\ }\bibfield  {title} {\enquote
  {\bibinfo {title} {Experimental investigation of far-field human cough
  airflows from healthy and influenza-infected subjects},}\ }\href {\doibase
  10.1111/ina.12680} {\bibfield  {journal} {\bibinfo  {journal} {Indoor Air}\
  ,\ \bibinfo {pages} {12}} (\bibinfo {year} {2020})}\BibitemShut {NoStop}%
\bibitem [{\citenamefont {Estevadeordal}\ and\ \citenamefont
  {Goss}()}]{Estevadeordal2005}%
  \BibitemOpen
  \bibfield  {author} {\bibinfo {author} {\bibnamefont {Estevadeordal},
  \bibfnamefont {J.}}\ and\ \bibinfo {author} {\bibnamefont {Goss},
  \bibfnamefont {L.}},\ }\bibfield  {title} {\enquote {\bibinfo {title} {{PIV
  with LED: particle shadow velocimetry (PSV) technique}},}\ }in\ \href@noop {}
  {\emph {\bibinfo {booktitle} {43rd AIAA aerospace sciences meeting and
  exhibit}}},\ p.~\bibinfo {pages} {37}\BibitemShut {NoStop}%
\bibitem [{\citenamefont {Frankel}\ \emph {et~al.}(2012)\citenamefont
  {Frankel}, \citenamefont {Bek{\"o}}, \citenamefont {Timm}, \citenamefont
  {Gustavsen}, \citenamefont {Hansen},\ and\ \citenamefont
  {Madsen}}]{Frankel2012}%
  \BibitemOpen
  \bibfield  {author} {\bibinfo {author} {\bibnamefont {Frankel}, \bibfnamefont
  {M.}}, \bibinfo {author} {\bibnamefont {Bek{\"o}}, \bibfnamefont {G.}},
  \bibinfo {author} {\bibnamefont {Timm}, \bibfnamefont {M.}}, \bibinfo
  {author} {\bibnamefont {Gustavsen}, \bibfnamefont {S.}}, \bibinfo {author}
  {\bibnamefont {Hansen}, \bibfnamefont {E.~W.}}, \ and\ \bibinfo {author}
  {\bibnamefont {Madsen}, \bibfnamefont {A.~M.}},\ }\bibfield  {title}
  {\enquote {\bibinfo {title} {{Seasonal Variations of Indoor Microbial
  Exposures and Their Relation to Temperature, Relative Humidity, and Air
  Exchange Rate}},}\ }\href {\doibase 10.1128/aem.02069-12} {\bibfield
  {journal} {\bibinfo  {journal} {Applied and Environmental Microbiology}\
  }\textbf {\bibinfo {volume} {78}},\ \bibinfo {pages} {8289--8297} (\bibinfo
  {year} {2012})}\BibitemShut {NoStop}%
\bibitem [{\citenamefont {Fuchs}, \citenamefont {Hain},\ and\ \citenamefont
  {K{\"a}hler}(2017)}]{Fuchs2017}%
  \BibitemOpen
  \bibfield  {author} {\bibinfo {author} {\bibnamefont {Fuchs}, \bibfnamefont
  {T.}}, \bibinfo {author} {\bibnamefont {Hain}, \bibfnamefont {R.}}, \ and\
  \bibinfo {author} {\bibnamefont {K{\"a}hler}, \bibfnamefont {C.~J.}},\
  }\bibfield  {title} {\enquote {\bibinfo {title} {{Non-iterative double-frame
  2D/3D particle tracking velocimetry}},}\ }\href {\doibase
  10.1007/s00348-017-2404-0} {\bibfield  {journal} {\bibinfo  {journal}
  {Experiments in Fluids}\ }\textbf {\bibinfo {volume} {58}},\ \bibinfo {pages}
  {119} (\bibinfo {year} {2017})}\BibitemShut {NoStop}%
\bibitem [{\citenamefont {Gong}\ \emph {et~al.}(2006)\citenamefont {Gong},
  \citenamefont {Tham}, \citenamefont {Melikov}, \citenamefont {Wyon},
  \citenamefont {Sekhar},\ and\ \citenamefont {Cheong}}]{Gong2006}%
  \BibitemOpen
  \bibfield  {author} {\bibinfo {author} {\bibnamefont {Gong}, \bibfnamefont
  {N.}}, \bibinfo {author} {\bibnamefont {Tham}, \bibfnamefont {K.~W.}},
  \bibinfo {author} {\bibnamefont {Melikov}, \bibfnamefont {A.~K.}}, \bibinfo
  {author} {\bibnamefont {Wyon}, \bibfnamefont {D.~P.}}, \bibinfo {author}
  {\bibnamefont {Sekhar}, \bibfnamefont {S.~C.}}, \ and\ \bibinfo {author}
  {\bibnamefont {Cheong}, \bibfnamefont {K.~W.}},\ }\bibfield  {title}
  {\enquote {\bibinfo {title} {The acceptable air velocity range for local air
  movement in the tropics},}\ }\href {\doibase 10.1080/10789669.2006.10391451}
  {\bibfield  {journal} {\bibinfo  {journal} {HVAC{\&}R Research}\ }\textbf
  {\bibinfo {volume} {12}},\ \bibinfo {pages} {1065--1076} (\bibinfo {year}
  {2006})}\BibitemShut {NoStop}%
\bibitem [{\citenamefont {Gupta}, \citenamefont {Lin},\ and\ \citenamefont
  {Chen}(2009)}]{Gupta2009}%
  \BibitemOpen
  \bibfield  {author} {\bibinfo {author} {\bibnamefont {Gupta}, \bibfnamefont
  {J.~K.}}, \bibinfo {author} {\bibnamefont {Lin}, \bibfnamefont {C.~H.}}, \
  and\ \bibinfo {author} {\bibnamefont {Chen}, \bibfnamefont {Q.}},\ }\bibfield
   {title} {\enquote {\bibinfo {title} {Flow dynamics and characterization of a
  cough},}\ }\href {\doibase 10.1111/j.1600-0668.2009.00619.x} {\bibfield
  {journal} {\bibinfo  {journal} {Indoor Air}\ }\textbf {\bibinfo {volume}
  {19}},\ \bibinfo {pages} {517--525} (\bibinfo {year} {2009})}\BibitemShut
  {NoStop}%
\bibitem [{\citenamefont {He}, \citenamefont {Jin},\ and\ \citenamefont
  {Yang}(2017)}]{HeGW2017}%
  \BibitemOpen
  \bibfield  {author} {\bibinfo {author} {\bibnamefont {He}, \bibfnamefont
  {G.~W.}}, \bibinfo {author} {\bibnamefont {Jin}, \bibfnamefont {G.~D.}}, \
  and\ \bibinfo {author} {\bibnamefont {Yang}, \bibfnamefont {Y.}},\ }\bibfield
   {title} {\enquote {\bibinfo {title} {{Space-Time Correlations and Dynamic
  Coupling in Turbulent Flows}},}\ }\href {\doibase
  doi:10.1146/annurev-fluid-010816-060309} {\bibfield  {journal} {\bibinfo
  {journal} {Annual Review of Fluid Mechanics}\ }\textbf {\bibinfo {volume}
  {49}},\ \bibinfo {pages} {51--70} (\bibinfo {year} {2017})}\BibitemShut
  {NoStop}%
\bibitem [{\citenamefont {He}, \citenamefont {Rubinstein},\ and\ \citenamefont
  {Wang}(2002)}]{HeGW2002}%
  \BibitemOpen
  \bibfield  {author} {\bibinfo {author} {\bibnamefont {He}, \bibfnamefont
  {G.~W.}}, \bibinfo {author} {\bibnamefont {Rubinstein}, \bibfnamefont {R.}},
  \ and\ \bibinfo {author} {\bibnamefont {Wang}, \bibfnamefont {L.~P.}},\
  }\bibfield  {title} {\enquote {\bibinfo {title} {Effects of subgrid-scale
  modeling on time correlations in large eddy simulation},}\ }\href {\doibase
  10.1063/1.1483877} {\bibfield  {journal} {\bibinfo  {journal} {Physics of
  Fluids}\ }\textbf {\bibinfo {volume} {14}},\ \bibinfo {pages} {2186--2193}
  (\bibinfo {year} {2002})}\BibitemShut {NoStop}%
\bibitem [{\citenamefont {Hessenkemper}\ and\ \citenamefont
  {Ziegenhein}(2018)}]{Hessenkemper2018}%
  \BibitemOpen
  \bibfield  {author} {\bibinfo {author} {\bibnamefont {Hessenkemper},
  \bibfnamefont {H.}}\ and\ \bibinfo {author} {\bibnamefont {Ziegenhein},
  \bibfnamefont {T.}},\ }\bibfield  {title} {\enquote {\bibinfo {title}
  {{Particle Shadow Velocimetry (PSV) in bubbly flows}},}\ }\href {\doibase
  https://doi.org/10.1016/j.ijmultiphaseflow.2018.04.015} {\bibfield  {journal}
  {\bibinfo  {journal} {International Journal of Multiphase Flow}\ }\textbf
  {\bibinfo {volume} {106}},\ \bibinfo {pages} {268--279} (\bibinfo {year}
  {2018})}\BibitemShut {NoStop}%
\bibitem [{\citenamefont {Holterman}(2003)}]{Holterman2003}%
  \BibitemOpen
  \bibfield  {author} {\bibinfo {author} {\bibnamefont {Holterman},
  \bibfnamefont {H.}},\ }\href@noop {} {\emph {\bibinfo {title} {Kinetics and
  evaporation of water drops in air}}},\ Vol.\ \bibinfo {volume} {2012}\
  (\bibinfo  {publisher} {Citeseer},\ \bibinfo {year} {2003})\BibitemShut
  {NoStop}%
\bibitem [{\citenamefont {Huang}\ \emph {et~al.}(2020)\citenamefont {Huang},
  \citenamefont {Fan}, \citenamefont {Li}, \citenamefont {Nie}, \citenamefont
  {Wang}, \citenamefont {Wang}, \citenamefont {Wang}, \citenamefont {Xia},
  \citenamefont {Zheng}, \citenamefont {Zuo},\ and\ \citenamefont
  {Huang}}]{Huang2020}%
  \BibitemOpen
  \bibfield  {author} {\bibinfo {author} {\bibnamefont {Huang}, \bibfnamefont
  {H.}}, \bibinfo {author} {\bibnamefont {Fan}, \bibfnamefont {C.}}, \bibinfo
  {author} {\bibnamefont {Li}, \bibfnamefont {M.}}, \bibinfo {author}
  {\bibnamefont {Nie}, \bibfnamefont {H.~L.}}, \bibinfo {author} {\bibnamefont
  {Wang}, \bibfnamefont {F.~B.}}, \bibinfo {author} {\bibnamefont {Wang},
  \bibfnamefont {H.}}, \bibinfo {author} {\bibnamefont {Wang}, \bibfnamefont
  {R.}}, \bibinfo {author} {\bibnamefont {Xia}, \bibfnamefont {J.}}, \bibinfo
  {author} {\bibnamefont {Zheng}, \bibfnamefont {X.}}, \bibinfo {author}
  {\bibnamefont {Zuo}, \bibfnamefont {X.}}, \ and\ \bibinfo {author}
  {\bibnamefont {Huang}, \bibfnamefont {J.}},\ }\bibfield  {title} {\enquote
  {\bibinfo {title} {{COVID-19: A Call for Physical Scientists and
  Engineers}},}\ }\href {\doibase 10.1021/acsnano.0c02618} {\bibfield
  {journal} {\bibinfo  {journal} {ACS Nano}\ }\textbf {\bibinfo {volume}
  {14}},\ \bibinfo {pages} {3747--3754} (\bibinfo {year} {2020})}\BibitemShut
  {NoStop}%
\bibitem [{\citenamefont {K{\"{a}}hler}\ and\ \citenamefont
  {Hain}(2020)}]{Kahler2020}%
  \BibitemOpen
  \bibfield  {author} {\bibinfo {author} {\bibnamefont {K{\"{a}}hler},
  \bibfnamefont {C.~J.}}\ and\ \bibinfo {author} {\bibnamefont {Hain},
  \bibfnamefont {R.}},\ }\bibfield  {title} {\enquote {\bibinfo {title}
  {Fundamental protective mechanisms of face masks against droplet
  infections},}\ }\href@noop {} {\bibfield  {journal} {\bibinfo  {journal}
  {Journal of Aerosol Science}\ ,\ \bibinfo {pages} {105617}} (\bibinfo {year}
  {2020})}\BibitemShut {NoStop}%
\bibitem [{\citenamefont {Krishnan}\ and\ \citenamefont
  {Mohseni}(2010)}]{Krishnan2010}%
  \BibitemOpen
  \bibfield  {author} {\bibinfo {author} {\bibnamefont {Krishnan},
  \bibfnamefont {G.}}\ and\ \bibinfo {author} {\bibnamefont {Mohseni},
  \bibfnamefont {K.}},\ }\bibfield  {title} {\enquote {\bibinfo {title} {An
  experimental study of a radial wall jet formed by the normal impingement of a
  round synthetic jet},}\ }\href {\doibase 10.1016/j.euromechflu.2010.03.001}
  {\bibfield  {journal} {\bibinfo  {journal} {European Journal of Mechanics
  B-Fluids}\ }\textbf {\bibinfo {volume} {29}},\ \bibinfo {pages} {269--277}
  (\bibinfo {year} {2010})}\BibitemShut {NoStop}%
\bibitem [{\citenamefont {Kukkonen}, \citenamefont {Vesala},\ and\
  \citenamefont {Kulmala}(1989)}]{Kukkonen1989}%
  \BibitemOpen
  \bibfield  {author} {\bibinfo {author} {\bibnamefont {Kukkonen},
  \bibfnamefont {J.}}, \bibinfo {author} {\bibnamefont {Vesala}, \bibfnamefont
  {T.}}, \ and\ \bibinfo {author} {\bibnamefont {Kulmala}, \bibfnamefont
  {M.}},\ }\bibfield  {title} {\enquote {\bibinfo {title} {{The Interdependence
  of Evaporation and Settling for Airborne Freely Falling Droplets}},}\ }\href
  {\doibase Doi 10.1016/0021-8502(89)90087-6} {\bibfield  {journal} {\bibinfo
  {journal} {Journal of Aerosol Science}\ }\textbf {\bibinfo {volume} {20}},\
  \bibinfo {pages} {749--763} (\bibinfo {year} {1989})}\BibitemShut {NoStop}%
\bibitem [{\citenamefont {Kwon}\ \emph {et~al.}(2012)\citenamefont {Kwon},
  \citenamefont {Park}, \citenamefont {Jang}, \citenamefont {Cho},
  \citenamefont {Park}, \citenamefont {Kim}, \citenamefont {Bae},\ and\
  \citenamefont {Jang}}]{Kwon2012}%
  \BibitemOpen
  \bibfield  {author} {\bibinfo {author} {\bibnamefont {Kwon}, \bibfnamefont
  {S.~B.}}, \bibinfo {author} {\bibnamefont {Park}, \bibfnamefont {J.}},
  \bibinfo {author} {\bibnamefont {Jang}, \bibfnamefont {J.}}, \bibinfo
  {author} {\bibnamefont {Cho}, \bibfnamefont {Y.}}, \bibinfo {author}
  {\bibnamefont {Park}, \bibfnamefont {D.~S.}}, \bibinfo {author} {\bibnamefont
  {Kim}, \bibfnamefont {C.}}, \bibinfo {author} {\bibnamefont {Bae},
  \bibfnamefont {G.~N.}}, \ and\ \bibinfo {author} {\bibnamefont {Jang},
  \bibfnamefont {A.}},\ }\bibfield  {title} {\enquote {\bibinfo {title} {Study
  on the initial velocity distribution of exhaled air from coughing and
  speaking},}\ }\href {\doibase 10.1016/j.chemosphere.2012.01.032} {\bibfield
  {journal} {\bibinfo  {journal} {Chemosphere}\ }\textbf {\bibinfo {volume}
  {87}},\ \bibinfo {pages} {1260--1264} (\bibinfo {year} {2012})}\BibitemShut
  {NoStop}%
\bibitem [{\citenamefont {Liu}\ \emph {et~al.}(2020)\citenamefont {Liu},
  \citenamefont {Ning}, \citenamefont {Chen}, \citenamefont {Guo},
  \citenamefont {Liu}, \citenamefont {Gali}, \citenamefont {Sun}, \citenamefont
  {Duan}, \citenamefont {Cai}, \citenamefont {Westerdahl}, \citenamefont {Liu},
  \citenamefont {Xu}, \citenamefont {Ho}, \citenamefont {Kan}, \citenamefont
  {Fu},\ and\ \citenamefont {Lan}}]{Liu2020}%
  \BibitemOpen
  \bibfield  {author} {\bibinfo {author} {\bibnamefont {Liu}, \bibfnamefont
  {Y.}}, \bibinfo {author} {\bibnamefont {Ning}, \bibfnamefont {Z.}}, \bibinfo
  {author} {\bibnamefont {Chen}, \bibfnamefont {Y.}}, \bibinfo {author}
  {\bibnamefont {Guo}, \bibfnamefont {M.}}, \bibinfo {author} {\bibnamefont
  {Liu}, \bibfnamefont {Y.}}, \bibinfo {author} {\bibnamefont {Gali},
  \bibfnamefont {N.~K.}}, \bibinfo {author} {\bibnamefont {Sun}, \bibfnamefont
  {L.}}, \bibinfo {author} {\bibnamefont {Duan}, \bibfnamefont {Y.}}, \bibinfo
  {author} {\bibnamefont {Cai}, \bibfnamefont {J.}}, \bibinfo {author}
  {\bibnamefont {Westerdahl}, \bibfnamefont {D.}}, \bibinfo {author}
  {\bibnamefont {Liu}, \bibfnamefont {X.}}, \bibinfo {author} {\bibnamefont
  {Xu}, \bibfnamefont {K.}}, \bibinfo {author} {\bibnamefont {Ho},
  \bibfnamefont {K.-f.}}, \bibinfo {author} {\bibnamefont {Kan}, \bibfnamefont
  {H.}}, \bibinfo {author} {\bibnamefont {Fu}, \bibfnamefont {Q.}}, \ and\
  \bibinfo {author} {\bibnamefont {Lan}, \bibfnamefont {K.}},\ }\bibfield
  {title} {\enquote {\bibinfo {title} {{Aerodynamic analysis of SARS-CoV-2 in
  two Wuhan hospitals}},}\ }\href {\doibase 10.1038/s41586-020-2271-3}
  {\bibfield  {journal} {\bibinfo  {journal} {Nature}\ }\textbf {\bibinfo
  {volume} {582}},\ \bibinfo {pages} {557--560} (\bibinfo {year}
  {2020})}\BibitemShut {NoStop}%
\bibitem [{\citenamefont {Matthews}\ \emph {et~al.}(1989)\citenamefont
  {Matthews}, \citenamefont {Thompson}, \citenamefont {Wilson}, \citenamefont
  {Hawthorne},\ and\ \citenamefont {Mage}}]{Matthews1989}%
  \BibitemOpen
  \bibfield  {author} {\bibinfo {author} {\bibnamefont {Matthews},
  \bibfnamefont {T.}}, \bibinfo {author} {\bibnamefont {Thompson},
  \bibfnamefont {C.}}, \bibinfo {author} {\bibnamefont {Wilson}, \bibfnamefont
  {D.}}, \bibinfo {author} {\bibnamefont {Hawthorne}, \bibfnamefont {A.}}, \
  and\ \bibinfo {author} {\bibnamefont {Mage}, \bibfnamefont {D.}},\ }\bibfield
   {title} {\enquote {\bibinfo {title} {Air velocities inside domestic
  environments: an important parameter in the study of indoor air quality and
  climate},}\ }\href@noop {} {\bibfield  {journal} {\bibinfo  {journal}
  {Environment International}\ }\textbf {\bibinfo {volume} {15}},\ \bibinfo
  {pages} {545--550} (\bibinfo {year} {1989})}\BibitemShut {NoStop}%
\bibitem [{\citenamefont {Mittal}, \citenamefont {Ni},\ and\ \citenamefont
  {Seo}(2020)}]{Mittal2020}%
  \BibitemOpen
  \bibfield  {author} {\bibinfo {author} {\bibnamefont {Mittal}, \bibfnamefont
  {R.}}, \bibinfo {author} {\bibnamefont {Ni}, \bibfnamefont {R.}}, \ and\
  \bibinfo {author} {\bibnamefont {Seo}, \bibfnamefont {J.~H.}},\ }\bibfield
  {title} {\enquote {\bibinfo {title} {{The flow physics of COVID-19}},}\
  }\href {\doibase 10.1017/jfm.2020.330} {\bibfield  {journal} {\bibinfo
  {journal} {Journal of Fluid Mechanics}\ }\textbf {\bibinfo {volume} {894}},\
  \bibinfo {pages} {14} (\bibinfo {year} {2020})}\BibitemShut {NoStop}%
\bibitem [{\citenamefont {Morawska}\ and\ \citenamefont
  {Milton}(2020)}]{Morawska2020}%
  \BibitemOpen
  \bibfield  {author} {\bibinfo {author} {\bibnamefont {Morawska},
  \bibfnamefont {L.}}\ and\ \bibinfo {author} {\bibnamefont {Milton},
  \bibfnamefont {D.~K.}},\ }\bibfield  {title} {\enquote {\bibinfo {title} {{It
  is Time to Address Airborne Transmission of COVID-19}},}\ }\href {\doibase
  10.1093/cid/ciaa939} {\bibfield  {journal} {\bibinfo  {journal} {Clinical
  Infectious Diseases}\ } (\bibinfo {year} {2020}),\
  10.1093/cid/ciaa939}\BibitemShut {NoStop}%
\bibitem [{\citenamefont {Philip}\ and\ \citenamefont
  {Marusic}(2012)}]{Philip2012}%
  \BibitemOpen
  \bibfield  {author} {\bibinfo {author} {\bibnamefont {Philip}, \bibfnamefont
  {J.}}\ and\ \bibinfo {author} {\bibnamefont {Marusic}, \bibfnamefont {I.}},\
  }\bibfield  {title} {\enquote {\bibinfo {title} {Large-scale eddies and their
  role in entrainment in turbulent jets and wakes},}\ }\href {\doibase
  10.1063/1.4719156} {\bibfield  {journal} {\bibinfo  {journal} {Physics of
  Fluids}\ }\textbf {\bibinfo {volume} {24}},\ \bibinfo {pages} {055108}
  (\bibinfo {year} {2012})}\BibitemShut {NoStop}%
\bibitem [{\citenamefont {Sahu}\ \emph {et~al.}(2013)\citenamefont {Sahu},
  \citenamefont {Tiwari}, \citenamefont {Bhangare},\ and\ \citenamefont
  {Pandit}}]{Sahu2013}%
  \BibitemOpen
  \bibfield  {author} {\bibinfo {author} {\bibnamefont {Sahu}, \bibfnamefont
  {S.~K.}}, \bibinfo {author} {\bibnamefont {Tiwari}, \bibfnamefont {M.}},
  \bibinfo {author} {\bibnamefont {Bhangare}, \bibfnamefont {R.~C.}}, \ and\
  \bibinfo {author} {\bibnamefont {Pandit}, \bibfnamefont {G.~G.}},\ }\bibfield
   {title} {\enquote {\bibinfo {title} {Particle size distribution of
  mainstream and exhaled cigarette smoke and predictive deposition in human
  respiratory tract},}\ }\href {\doibase 10.4209/aaqr.2012.02.0041} {\bibfield
  {journal} {\bibinfo  {journal} {Aerosol and Air Quality Research}\ }\textbf
  {\bibinfo {volume} {13}},\ \bibinfo {pages} {324--332} (\bibinfo {year}
  {2013})}\BibitemShut {NoStop}%
\bibitem [{\citenamefont {Scarano}(2002)}]{Scarano2002}%
  \BibitemOpen
  \bibfield  {author} {\bibinfo {author} {\bibnamefont {Scarano}, \bibfnamefont
  {F.}},\ }\bibfield  {title} {\enquote {\bibinfo {title} {{Iterative image
  deformation methods in PIV}},}\ }\href {\doibase 10.1088/0957-0233/13/1/201}
  {\bibfield  {journal} {\bibinfo  {journal} {Measurement Science and
  Technology}\ }\textbf {\bibinfo {volume} {13}},\ \bibinfo {pages} {R1--R19}
  (\bibinfo {year} {2002})}\BibitemShut {NoStop}%
\bibitem [{\citenamefont {Scharfman}\ \emph {et~al.}(2016)\citenamefont
  {Scharfman}, \citenamefont {Techet}, \citenamefont {Bush},\ and\
  \citenamefont {Bourouiba}}]{Scharfman2016}%
  \BibitemOpen
  \bibfield  {author} {\bibinfo {author} {\bibnamefont {Scharfman},
  \bibfnamefont {B.~E.}}, \bibinfo {author} {\bibnamefont {Techet},
  \bibfnamefont {A.~H.}}, \bibinfo {author} {\bibnamefont {Bush}, \bibfnamefont
  {J.~W.~M.}}, \ and\ \bibinfo {author} {\bibnamefont {Bourouiba},
  \bibfnamefont {L.}},\ }\bibfield  {title} {\enquote {\bibinfo {title}
  {Visualization of sneeze ejecta: steps of fluid fragmentation leading to
  respiratory droplets},}\ }\href {\doibase 10.1007/s00348-015-2078-4}
  {\bibfield  {journal} {\bibinfo  {journal} {Experiments in Fluids}\ }\textbf
  {\bibinfo {volume} {57}},\ \bibinfo {pages} {24} (\bibinfo {year}
  {2016})}\BibitemShut {NoStop}%
\bibitem [{\citenamefont {Tang}\ \emph {et~al.}(2009)\citenamefont {Tang},
  \citenamefont {Liebner}, \citenamefont {Craven},\ and\ \citenamefont
  {Settles}}]{Tang2009}%
  \BibitemOpen
  \bibfield  {author} {\bibinfo {author} {\bibnamefont {Tang}, \bibfnamefont
  {J.~W.}}, \bibinfo {author} {\bibnamefont {Liebner}, \bibfnamefont {T.~J.}},
  \bibinfo {author} {\bibnamefont {Craven}, \bibfnamefont {B.~A.}}, \ and\
  \bibinfo {author} {\bibnamefont {Settles}, \bibfnamefont {G.~S.}},\
  }\bibfield  {title} {\enquote {\bibinfo {title} {A schlieren optical study of
  the human cough with and without wearing masks for aerosol infection
  control},}\ }\href {\doibase 10.1098/rsif.2009.0295.focus} {\bibfield
  {journal} {\bibinfo  {journal} {J R Soc Interface}\ }\textbf {\bibinfo
  {volume} {6 Suppl 6}},\ \bibinfo {pages} {S727--36} (\bibinfo {year}
  {2009})}\BibitemShut {NoStop}%
\bibitem [{\citenamefont {Verma}, \citenamefont {Dhanak},\ and\ \citenamefont
  {Frankenfield}(2020)}]{Verma2020}%
  \BibitemOpen
  \bibfield  {author} {\bibinfo {author} {\bibnamefont {Verma}, \bibfnamefont
  {S.}}, \bibinfo {author} {\bibnamefont {Dhanak}, \bibfnamefont {M.}}, \ and\
  \bibinfo {author} {\bibnamefont {Frankenfield}, \bibfnamefont {J.}},\
  }\bibfield  {title} {\enquote {\bibinfo {title} {Visualizing the
  effectiveness of face masks in obstructing respiratory jets},}\ }\href
  {\doibase 10.1063/5.0016018} {\bibfield  {journal} {\bibinfo  {journal}
  {Physics of Fluids}\ }\textbf {\bibinfo {volume} {32}},\ \bibinfo {pages}
  {061708} (\bibinfo {year} {2020})}\BibitemShut {NoStop}%
\bibitem [{\citenamefont {Wells}(1955)}]{Wells1955}%
  \BibitemOpen
  \bibfield  {author} {\bibinfo {author} {\bibnamefont {Wells}, \bibfnamefont
  {W.~F.}},\ }\bibfield  {title} {\enquote {\bibinfo {title} {Airborne
  contagion and air hygiene. an ecological study of droplet infections},}\
  }\href@noop {} {\bibfield  {journal} {\bibinfo  {journal} {Airborne Contagion
  and Air Hygiene. An Ecological Study of Droplet Infections.}\ } (\bibinfo
  {year} {1955})}\BibitemShut {NoStop}%
\bibitem [{\citenamefont {Westerweel}\ and\ \citenamefont
  {Scarano}(2005)}]{Westerweel2005}%
  \BibitemOpen
  \bibfield  {author} {\bibinfo {author} {\bibnamefont {Westerweel},
  \bibfnamefont {J.}}\ and\ \bibinfo {author} {\bibnamefont {Scarano},
  \bibfnamefont {F.}},\ }\bibfield  {title} {\enquote {\bibinfo {title}
  {{Universal outlier detection for PIV data}},}\ }\href {\doibase
  10.1007/s00348-005-0016-6} {\bibfield  {journal} {\bibinfo  {journal}
  {Experiments in Fluids}\ }\textbf {\bibinfo {volume} {39}},\ \bibinfo {pages}
  {1096--1100} (\bibinfo {year} {2005})}\BibitemShut {NoStop}%
\bibitem [{\citenamefont {White}\ and\ \citenamefont
  {Corfield}(2006)}]{White2006}%
  \BibitemOpen
  \bibfield  {author} {\bibinfo {author} {\bibnamefont {White}, \bibfnamefont
  {F.~M.}}\ and\ \bibinfo {author} {\bibnamefont {Corfield}, \bibfnamefont
  {I.}},\ }\href@noop {} {\emph {\bibinfo {title} {Viscous fluid flow}}},\
  Vol.~\bibinfo {volume} {3}\ (\bibinfo  {publisher} {McGraw-Hill New York},\
  \bibinfo {year} {2006})\BibitemShut {NoStop}%
\bibitem [{\citenamefont {Xie}\ \emph {et~al.}(2007)\citenamefont {Xie},
  \citenamefont {Li}, \citenamefont {Chwang}, \citenamefont {Ho},\ and\
  \citenamefont {Seto}}]{Xie2007}%
  \BibitemOpen
  \bibfield  {author} {\bibinfo {author} {\bibnamefont {Xie}, \bibfnamefont
  {X.}}, \bibinfo {author} {\bibnamefont {Li}, \bibfnamefont {Y.}}, \bibinfo
  {author} {\bibnamefont {Chwang}, \bibfnamefont {A.~T.~Y.}}, \bibinfo {author}
  {\bibnamefont {Ho}, \bibfnamefont {P.~L.}}, \ and\ \bibinfo {author}
  {\bibnamefont {Seto}, \bibfnamefont {W.~H.}},\ }\bibfield  {title} {\enquote
  {\bibinfo {title} {{How far droplets can move in indoor environments -
  revisiting the Wells evaporation-falling curve}},}\ }\href {\doibase
  10.1111/j.1600-0668.2007.00469.x} {\bibfield  {journal} {\bibinfo  {journal}
  {Indoor Air}\ }\textbf {\bibinfo {volume} {17}},\ \bibinfo {pages} {211--225}
  (\bibinfo {year} {2007})}\BibitemShut {NoStop}%
\bibitem [{\citenamefont {Xie}\ \emph {et~al.}(2009)\citenamefont {Xie},
  \citenamefont {Li}, \citenamefont {Sun},\ and\ \citenamefont
  {Liu}}]{Xie2009}%
  \BibitemOpen
  \bibfield  {author} {\bibinfo {author} {\bibnamefont {Xie}, \bibfnamefont
  {X.}}, \bibinfo {author} {\bibnamefont {Li}, \bibfnamefont {Y.}}, \bibinfo
  {author} {\bibnamefont {Sun}, \bibfnamefont {H.}}, \ and\ \bibinfo {author}
  {\bibnamefont {Liu}, \bibfnamefont {L.}},\ }\bibfield  {title} {\enquote
  {\bibinfo {title} {Exhaled droplets due to talking and coughing},}\ }\href
  {\doibase 10.1098/rsif.2009.0388.focus} {\bibfield  {journal} {\bibinfo
  {journal} {Journal of the Royal Society, Interface}\ }\textbf {\bibinfo
  {volume} {6 Suppl 6}},\ \bibinfo {pages} {S703--S714} (\bibinfo {year}
  {2009})}\BibitemShut {NoStop}%
\bibitem [{\citenamefont {Xu}, \citenamefont {Feng},\ and\ \citenamefont
  {Wang}(2013)}]{Xu2013}%
  \BibitemOpen
  \bibfield  {author} {\bibinfo {author} {\bibnamefont {Xu}, \bibfnamefont
  {Y.}}, \bibinfo {author} {\bibnamefont {Feng}, \bibfnamefont {L.~H.}}, \ and\
  \bibinfo {author} {\bibnamefont {Wang}, \bibfnamefont {J.~J.}},\ }\bibfield
  {title} {\enquote {\bibinfo {title} {Experimental investigation of a
  synthetic jet impinging on a fixed wall},}\ }\href {\doibase ARTN 1512
  10.1007/s00348-013-1512-8} {\bibfield  {journal} {\bibinfo  {journal}
  {Experiments in Fluids}\ }\textbf {\bibinfo {volume} {54}} (\bibinfo {year}
  {2013}),\ ARTN 1512 10.1007/s00348-013-1512-8}\BibitemShut {NoStop}%
\bibitem [{\citenamefont {Xu}\ \emph {et~al.}(2018)\citenamefont {Xu},
  \citenamefont {Wang}, \citenamefont {Feng}, \citenamefont {He},\ and\
  \citenamefont {Wang}}]{Xu2018}%
  \BibitemOpen
  \bibfield  {author} {\bibinfo {author} {\bibnamefont {Xu}, \bibfnamefont
  {Y.}}, \bibinfo {author} {\bibnamefont {Wang}, \bibfnamefont {J.~J.}},
  \bibinfo {author} {\bibnamefont {Feng}, \bibfnamefont {L.~H.}}, \bibinfo
  {author} {\bibnamefont {He}, \bibfnamefont {G.~S.}}, \ and\ \bibinfo {author}
  {\bibnamefont {Wang}, \bibfnamefont {Z.~Y.}},\ }\bibfield  {title} {\enquote
  {\bibinfo {title} {Laminar vortex rings impinging onto porous walls with a
  constant porosity},}\ }\href {\doibase 10.1017/jfm.2017.878} {\bibfield
  {journal} {\bibinfo  {journal} {Journal of Fluid Mechanics}\ }\textbf
  {\bibinfo {volume} {837}},\ \bibinfo {pages} {729--764} (\bibinfo {year}
  {2018})}\BibitemShut {NoStop}%
\bibitem [{\citenamefont {Zhang}\ and\ \citenamefont {Wang}(2007)}]{Zhang2007}%
  \BibitemOpen
  \bibfield  {author} {\bibinfo {author} {\bibnamefont {Zhang}, \bibfnamefont
  {P.~F.}}\ and\ \bibinfo {author} {\bibnamefont {Wang}, \bibfnamefont
  {J.~J.}},\ }\bibfield  {title} {\enquote {\bibinfo {title} {Novel signal wave
  pattern for efficient synthetic jet generation},}\ }\href {\doibase
  10.2514/1.25445} {\bibfield  {journal} {\bibinfo  {journal} {Aiaa Journal -
  AIAA J}\ }\textbf {\bibinfo {volume} {45}},\ \bibinfo {pages} {1058--1065}
  (\bibinfo {year} {2007})}\BibitemShut {NoStop}%
\bibitem [{\citenamefont {Zhao}\ and\ \citenamefont {He}(2009)}]{HeGW2009}%
  \BibitemOpen
  \bibfield  {author} {\bibinfo {author} {\bibnamefont {Zhao}, \bibfnamefont
  {X.}}\ and\ \bibinfo {author} {\bibnamefont {He}, \bibfnamefont {G.~W.}},\
  }\bibfield  {title} {\enquote {\bibinfo {title} {Space-time correlations of
  fluctuating velocities in turbulent shear flows},}\ }\href {\doibase
  10.1103/PhysRevE.79.046316} {\bibfield  {journal} {\bibinfo  {journal}
  {Physical Review E}\ }\textbf {\bibinfo {volume} {79}},\ \bibinfo {pages}
  {12} (\bibinfo {year} {2009})}\BibitemShut {NoStop}%
\bibitem [{\citenamefont {Zhu}, \citenamefont {Kato},\ and\ \citenamefont
  {Yang}(2006)}]{Zhu2006}%
  \BibitemOpen
  \bibfield  {author} {\bibinfo {author} {\bibnamefont {Zhu}, \bibfnamefont
  {S.}}, \bibinfo {author} {\bibnamefont {Kato}, \bibfnamefont {S.}}, \ and\
  \bibinfo {author} {\bibnamefont {Yang}, \bibfnamefont {J.-H.}},\ }\bibfield
  {title} {\enquote {\bibinfo {title} {Study on transport characteristics of
  saliva droplets produced by coughing in a calm indoor environment},}\ }\href
  {\doibase https://doi.org/10.1016/j.buildenv.2005.06.024} {\bibfield
  {journal} {\bibinfo  {journal} {Building and Environment}\ }\textbf {\bibinfo
  {volume} {41}},\ \bibinfo {pages} {1691--1702} (\bibinfo {year}
  {2006})}\BibitemShut {NoStop}%
\end{thebibliography}%

\end{document}